\documentclass[12pt]{iopart}
\usepackage{doi}

\usepackage{iopams}  
\usepackage{url}
\usepackage{amssymb}
\usepackage{amsfonts}

\usepackage{graphicx} 
\usepackage[graphicx]{realboxes} 
\usepackage{color}
\usepackage{xr}
\usepackage[utf8]{inputenc}
\usepackage{bm} 
\usepackage{comment} 
\usepackage{rotating} 
\usepackage{xspace}    

\usepackage{bm}
\usepackage{hyperref}
\usepackage{lipsum}
\hypersetup{breaklinks=true,colorlinks=true,linkcolor=blue,citecolor=blue,filecolor=magenta,urlcolor=cyan}

\usepackage{orcidlink}

\usepackage{cleveref} 
\crefname{equation}{equation}{equations}
\Crefname{equation}{Equation}{Equations}

\newcommand{\fandf}{FaNDF\xspace}
\newcommand{\ecnoise}{ECNoise\xspace}
\newcommand{\pounders}{POUNDerS\xspace}
\newcommand{\dataset}{\mathcal{D}}                  
\newcommand{\nd}{n} 
\newcommand{\nx}{p} 

\newcommand{\xbhat}{\hat{\xb}}

\newcommand{\xbhata}{\hat{\xb}_{13\mathrm{D}}}

\newcommand{\Jrhata}{\hat{J}_{13\mathrm{D}}}
\newcommand{\xbhatb}{\hat{\xb}_{14\mathrm{D}}}

\newcommand{\Jrhatb}{\hat{J}_{14\mathrm{D}}}

\newcommand{\htwovm}{h^{\mathrm{v}}_{2-}}
\newcommand{\delfxi}{f_{\mathrm{ex},-}^{\xi}}
 
\renewcommand{\vec}{\boldsymbol}

\newcommand{\nub}{\bm{\nu}}

\newcommand{\xb}{\vec{x}}        
                                                                      
\newcommand{\R}{\mathbb{R}} 

\newcommand{\sfrac}[2]{{\textstyle\frac{#1}{#2}}}
\newcommand{\half}{\textstyle\frac{1}{2}}

\newcommand{\cmmnt}[1]{\ignorespaces}

\definecolor{dgreen}{rgb}{0.0,0.6,0.0} 
\newcommand{\parm}[1]{{\color{magenta} #1}}
\newcommand{\fixp}[1]{{\color{dgreen} #1}}

\DeclareGraphicsExtensions{.pdf}
\graphicspath{{./images/}}

\begin{document}

\title[Fayans EDF optimization]{Extended Fayans energy density functional: optimization and analysis}
\author{Paul-Gerhard Reinhard\orcidlink{0000-0002-4505-1552}$^1$, Jared O'Neal\orcidlink{0000-0003-2603-7314}$^2$, Stefan M Wild\orcidlink{0000-0002-6099-2772}$^{3,4}$, Witold~Nazarewicz\orcidlink{0000-0002-8084-7425}$^{5,6}$}

\address{$^1$ Institut  f\"ur  Theoretische  Physik  II,  Universit\"at  Erlangen-N\"urnberg,  D-91058  Erlangen,  Germany}
\address{$^2$ Mathematics and Computer Science Division, Argonne National Laboratory, Lemont, IL 60439, USA}
\address{$^3$ Applied Mathematics and Computational Research Division, Lawrence Berkeley National Laboratory, Berkeley, California 94720, USA}
\address{$^4$
Department of Industrial Engineering and Management Sciences, Northwestern University, Evanston, IL 60208, USA}
\address{$^5$
Facility for Rare Isotope Beams, Michigan  State  University,  East  Lansing,  MI  48824,  USA}
\address{$^6$
Department  of  Physics  and  Astronomy, Michigan  State  University,  East  Lansing,  MI 48824,  USA}

\ead{joneal@anl.gov}

\begin{abstract}
The Fayans energy density functional (EDF) has been very successful in describing global nuclear properties (binding energies, charge radii, and especially differences of radii) within nuclear density functional theory.
In a recent study,
supervised machine learning methods were used 
to calibrate the  Fayans EDF. Building on this experience, in this work we  
explore the effect of adding isovector pairing terms, which are responsible for different proton and neutron pairing fields, by comparing a 13D model without the isovector pairing term against the extended 14D model. At the heart of the
calibration is a  carefully selected heterogeneous dataset of experimental observables representing ground-state properties of spherical even-even nuclei. To quantify the impact of the calibration dataset on model parameters and the importance of the new terms, we  carry out advanced sensitivity and correlation analysis on both models.
The extension to 14D improves the overall quality of the model by about 30\%. The enhanced degrees of freedom of the 14D model reduce correlations between model parameters and enhance sensitivity.
\end{abstract}

\vspace{2pc} 

\section{Introduction}

Nuclear density functional theory (DFT)  \cite{Bender2003,Duguet2014,Schunck2019a} is a quantum many-body method applicable across the whole nuclear landscape. At the heart of nuclear DFT lies the energy density functional (EDF) that represents an effective internucleon interaction. The EDF is a functional of various nucleonic densities and currents, which are  usually
assumed to be local. The EDF coupling constants are usually
adjusted to experimental data and---in many cases---to selected nuclear matter parameters. The   validated global  EDFs often provide a level of accuracy typical of phenomenological approaches based on parameters locally optimized to the experiment and enable extrapolations toward particle drip lines and beyond   \cite{Neufcourt2020a}.

The EDF developed by S.A.~Fayans and collaborators \cite{Fayans1994a,Fayans1998,Fayans2000,Tolokonnikov2015} turned out to be particularly useful since it 
was designed to describe the ground-state properties of finite nuclei. The volume part of the functional was adjusted to reproduce   the  microscopic equation of state of the nucleonic matter \cite{Fayans1998}.  In this sense the functional could be considered ``universal.''
By employing a density-dependent pairing
functional with gradient terms, the Fayans EDF
was able
 to explain the odd-even staggering effect in charge radii \cite{Fayans1994a,Fayans2000}.

In~\cite{Reinhard2017}, detailed analysis of the Fayans EDF was carried out. Various optimization strategies were explored to arrive 
at a consistent description of odd-even staggering of binding energies and charge radii. Next, the functional was extended
to weakly bound nuclei \cite{Miller2019} and long isotopic chains,
to that end invoking Hartree--Fock--Bogoliubov (HFB) pairing instead of the simpler 
 Bardeen–-Cooper–-Schrieffer (BCS) approach. These functionals were
subsequently used for the interpretation of experimental data on charge radii
\cite{Hammen2018,Gorges2019,deGroote2020,Yordanov2020,Borzov2020,Koszorus2021,Reponen2021,ReinhardM2022,Kortelainen2022,Malbrunot2022,Geldhof2022,Sommer2022,Hur2022,Konig2023}.

Recently, the Fayans functional was extended by allowing separate
pairing strengths for proton and neutrons, that is, pairing isovector terms. Indeed, in order to accommodate the experimental odd-even mass staggering, 
the effective pairing interaction in atomic nuclei requires larger strength in the proton pairing channel than in the neutron pairing channel
\cite{Bertsch2009}.
Such an extension enhances the
flexibility to accommodate the radius trends in isotopic chains also
in heavier nuclei; for a preliminary application see \cite{Karthein2023}.

Following the previous large-scale calibration studies of Skyrme EDFs
\cite{UNEDF0,UNEDF1,UNEDF2,McDonnellPRL15}, in~\cite{Bollapragada2021} various
supervised machine learning methods were employed 
to optimize  the Fayans EDF. Building on this experience, in this study we  
explore the effect of adding isovector pairing terms. This is done
based on the dataset of \cite{Reinhard2017}. 
We compare fits
with and without the pairing isovector terms and provide advanced sensitivity analysis of the resulting model.

\section{The Fayans functional}
\label{sec:Fayfunc}

The Fayans EDF is a nonrelativistic energy density functional
similar to the widely used Skyrme functional \cite{Bender2003}, but
with more flexibility in density dependence and pairing.  We use it
here in the form of the original FaNDF0 parameterization
\cite{Fayans1998}. The functional is formulated in terms of particle density
$\rho_t$, kinetic density $\tau_t$, spin-orbit current $\vec{J}_t$,
and pairing densities $\breve\rho_t$, where the isospin index $t$
labels isoscalar ($t= 0$) and isovector ($t = 1$) densities; for details see \ref{sec:dens}.  
The isoscalar and isovector densities can be expressed in terms of
proton ($p$, $\tau_3=-1$) and neutron ($n$, $\tau_3=+1$) densities, for example, 
\begin{equation}
  \rho_0
  =
  \rho_n+\rho_p,~~\rho_1=\rho_n-\rho_p,
\end{equation}
and similarly for the other densities.
It is convenient to use also the dimensionless densities
\begin{equation}
  x_t
  =
  \frac{\rho_t}{\rho_\mathrm{sat}}
  \;,\;
  x_\mathrm{pair}
  =
  \frac{\rho_0}{\rho_\mathrm{pair}}
  \;,
\end{equation}
where $\rho_\mathrm{sat}$ and ${\rho_\mathrm{pair}}$ are scaling parameters
of the Fayans EDF.

Within DFT,  the total energy of the system is given by $E=\int d^{3}r\mathcal{E}(\vec{r})$, where the local
energy density $\mathcal{E}$ is a functional of the local isoscalar and isovector particle and pairing densities and currents.  The energy density of the Fayans EDF
is composed from volume, surface, spin-orbit, and pairing
terms. We use it here in the following form:
\numparts
\begin{eqnarray}
  \mathcal{E}_{\mathrm{Fy}}
  &=&
  \mathcal{E}_{\mathrm{Fy}}^\mathrm{v}(\rho)
  +\mathcal{E}_{\mathrm{Fy}}^\mathrm{s}(\rho)
  +\mathcal{E}_{\mathrm{Fy}}^\mathrm{ls}(\rho,\vec{J})
  +\mathcal{E}_{\mathrm{Fy}}^\mathrm{pair}(\rho,\breve{\rho})
\label{EFay}
\\
  \mathcal{E}_{\mathrm{Fy}}^\mathrm{v}
  &=&
  \sfrac{1}{3}\varepsilon_F\rho_\mathrm{sat}
  \left[
  {a_+^\mathrm{v}}
  \frac{1\!-\!{h_{1+}^\mathrm{v}}x_0^{{\sigma}}}
       {1\!+\!{h_{2+}^\mathrm{v}}x_0^{{\sigma}}}x_0^2
  +
  {a_-^\mathrm{v}}
  \frac{1\!-\!{h_{1-}^\mathrm{v}}x_0}
       {1\!+\!{h_{2-}^\mathrm{v}}x_0}x_1^2
  \right]
\label{EFay-dens}
\\
  \mathcal{E}_\mathrm{Fy}^\mathrm{s}
  &=& 
  \sfrac{1}{3}\varepsilon_F\rho_\mathrm{sat}
  \frac{a_+^\mathrm{s}r_s^2(\vec{\nabla} x_0)^2}
       {1 
        +h_{\nabla}^\mathrm{s}r_s^2(\vec{\nabla} x_0)^2}
\label{EFay-grad} 
\\
  \mathcal{E}_\mathrm{Fy}^\mathrm{ls}
  &=&
  \frac{4\varepsilon_F r_s^2}{3\rho_\mathrm{sat}}
  \left(\kappa\rho_0\vec{\nabla}\cdot\vec{J}_0
  +
  \kappa'\rho_1\vec{\nabla}\cdot\vec{J}_1  
  +g \vec{J}_0^2 + g'\vec{J}_1^2
  \right)
\label{eq:EFy-ls}
\\
  \mathcal{E}_{\mathrm{Fy},q}^\mathrm{pair}
  &=&
  \frac{4}{\varepsilon_F}{3\rho_\mathrm{sat}}
 {\breve\rho_q}^2
  \left[f_{\mathrm{ex},+}^\xi-\tau_{3q}f_{\mathrm{ex},-}^\xi
       +h_{1+}^\xi x_\mathrm{pair}^{\gamma}
       +h_\nabla^\xi r_s^2 (\vec{\nabla} x_\mathrm{pair})^2\right].
\label{eq:ep2}
\end{eqnarray}
\endnumparts
Several EDF parameters are fixed a priori. These are
$\hbar^2/2m_p=20.749811$\,MeV\,fm$^2$,
$\hbar^2/2m_n=20.721249$\,MeV\,fm$^2$, $e^2=1.43996448$\,MeV\,fm,
$\rho_\mathrm{sat}=0.16$\,fm$^{-3}$,
$\rho_\mathrm{pair}=\rho_\mathrm{sat}$, $\sigma=1/3$, and $\gamma=2/3$.
The saturation density $\rho_\mathrm{sat}$ determines also the
auxiliary parameters Wigner--Seitz radius $r_s=
(3/4\pi\rho_\mathrm{sat})^{1/3}$ and Fermi energy
$\varepsilon_F=(9\pi/8)^{2/3}\hbar^2/2mr_s^2$. The saturation density
$\rho_\mathrm{sat}$ is a fixed scaling parameter, not identical to
the physical equilibrium density $\rho_\mathrm{eq}$ that is a result of
the model. Note the factor 4 in the pairing functional (\ref{eq:ep2});
the paper \cite{Reinhard2017} had a misprint at that place showing only a factor of 2.

Besides the Fayans nuclear energy $\mathcal{E}_\mathrm{Fy}$, the
total energy accounts also for Coulomb energy (direct and exchange) and
the center-of-mass correction term. These are standard terms without free
parameters \cite{Bender2003}, and hence they are not documented here.
The pairing functional is complemented by prescription for the cutoff in pairing space, which is explained in \ref{sec:dens}.
Altogether, the discussed Fayans model has $p=13 (14)$  free parameters: 
six in the volume term ($a_\pm^\mathrm{v},h_{1\pm}^\mathrm{v},h_{2\pm}^\mathrm{v}$),
two in the surface term ($a_+^\mathrm{s},h_{\nabla}^\mathrm{s}$),
two in the spin-orbit term ($\kappa,\kappa'$), and
three (four) in the pairing term 
($f_{\mathrm{ex},+}^\xi,[f_{\mathrm{ex},-}^\xi],h_+^\xi,h_{\nabla+}$).
Five of the six volume parameters can be expressed in terms of five
nuclear matter properties (NMPs), namely, equilibrium density
$\rho_\mathrm{eq}$, energy per nucleon $E_{\rm B}/A$, incompressibility
$K$, symmetry energy $J$, and slope of symmetry energy $L$; for their definition in terms of the energy functional see 
\ref{sec:NMP}.  There remains only $h_{2-}^\mathrm{v}$ as a
direct volume parameter.  This recoupling has the advantage that the
rather technical model parameters are replaced by more physical droplet
model constants. We use the parameters in this recoupled form.

The numerical treatment is based on the spherical Hartree--Fock code
 \cite{Rei91aR}.
The spherical DFT equations are solved on a numerical 1D radial grid 
with five-point finite differences for derivatives, a spacing of 0.3\,fm, and
a box size from 9.6 fm for light nuclei to 13.8 fm for heavy ones.
The solution is determined iteratively by using the accelerated gradient technique. For the BCS pairing cutoff, we use a soft cutoff with the Fermi profile \cite{Kri90a}; see \ref{sec:dens} for details.

A few words are in order about the numerical realization of computing nuclear
properties for the Fayans functional. The largest part of the
computations, namely, preparing the observables for the optimization routine
\pounders and subsequent analysis of the results, is done
with a spherical 1D code. The radial wavefunctions and fields are
represented on a spatial grid along radial directions. The ground state
is found by using accelerated gradient iterations on the energy
landscape. The numerical basics are explained in detail in
\cite{Rei91aR}.  In~\sref{sec:chains} we also analyze the
predictions for deformed nuclei along a selection of isotopic
chains. These embrace also deformed nuclei. The deformed calculations
are performed by using a cylindrical 2D grid in coordinate space. As in
the 1D case, accelerated gradient iteration coupled to the BCS
iterations is used to find the ground state. The 2D code, coined {\tt
  SkyAx}, is explained in detail in \cite{Reinhard2021}.

\section{Optimization and local analysis}
\label{sec:optimization}

\subsection{Problem definition: the objective function}

The \fandf DFT package uses the parameterized Fayans EDF to obtain the model value $m(\nub_i; \xb)$ of a given observable for a given nucleus, both specified by the input $\nub_i$, at a desired parameter-space point $\xb \in \R^{\nx}$.
For a particular dataset
$\dataset = \left\{(\nub_i, d_i)\right\}_{i=1}^{\nd}$
we construct the weighted least-squares objective function
\begin{equation}
\label{eqn:ObjFcn}
f(\xb; \dataset) = \sum_{i=1}^{\nd} \left(\frac{d_i - m(\nub_i; \xb)}{w_i}\right)^2 = \sum_{i=1}^{\nd} \delta_i^2(\xb; \dataset),
\end{equation}
where the $w_i$, which we refer to as ``adopted errors," are positive numbers discussed below and where $\delta_i$ are residuals.
Note that the residuals are dimensionless by virtue of the weights
$w_i$. This allows the accumulation of contributions from different physical
observables. Effectively, we deal with a dimensionless dataset
$\tilde\dataset = \left\{(\nub_i, \tilde{d}_i)\right\}_{i=1}^{\nd}$
with 
\begin{equation}\tilde{d}_i\equiv 
  \frac{d_i}{w_i},
\label{eq:dimlessdata}
\end{equation}
which allows us to compare variations of $\tilde{d}_i$ from physically
different types of observables \cite{Birge1932,DNR14}.

In this paper we use the iterative derivative-free
optimization software \pounders \cite{SWCHAP14} to approximate a nonlinear least-squares
local minimizer $\hat{\xb}$ associated with the dataset such that
\begin{equation}
\label{eqn:min_problem}
f(\hat{\xb}; \tilde\dataset) \approx \min_{\xb \in \R^p}\,f(\xb; \tilde\dataset).
\end{equation}

\subsection{Regression analysis}
\label{sec:Regression}

Optimization has reached its goal if an approximate minimum of the objective
function $f$ is found. In addition to the minimum point defining the optimal parameter set $\hat{\xb}$,  the behavior of $f$ around  
$\hat{\xb}$ carries useful information; the local behavior determines the response to
slightly varying conditions such as noise in the data. When endowed with a particular statistical interpretation, the local behavior 
determines a range of reasonable parameters by using $f$ as the
generator of a probability distribution in parameter space.  The profile of $f$, soft or
steep, determines the width of the probability distribution near $\hat{\xb}$. The
vicinity near  $\hat{\xb}$ can be described by a Taylor expansion with respect to $\xb$.
  The first derivative at a
minimum disappears; that is,  $\partial_{x_\alpha}f=0$. The second
derivative at a minimum can be approximated as
\begin{equation}
\partial_{x_\alpha}\partial_{x_\beta}f
\big|_{\hat{\xb}}
  \approx 
  C_{\alpha\beta}
  =
  \sum_i J_{i\alpha}J_{i\beta},
\label{eq:Jacobian}
\end{equation}
where $\hat{J}\equiv J_{i\alpha}=\partial_{x_\alpha}\delta_i$ is the Jacobian matrix and $\hat{C}\equiv C_{\alpha\beta}$ is merely shorthand
for an approximation of the second derivative of the objective function, which characterizes the leeway of the model
parameters.  In certain statistical settings, the inverse $\hat{C}^{-1}$ can be interpreted as proportional to an approximate covariance matrix. Its
diagonal elements give an estimate of the standard deviation $\hat{\sigma}_\alpha$ of parameter
$x_\alpha$ as
\begin{equation}
  \hat{\sigma}_\alpha
  =
  \hat{\sigma}\sqrt{(\hat{C}^{-1})_{\alpha\alpha}}
  \;,\;
  \hat{\sigma}^2
  =
  \frac{f(\hat{\xb})}{n-p}
  \;.
\label{eq:sigmaa}
\end{equation}
The value $\hat{\sigma}_\alpha$ sets a
natural scale for variations  of $x_\alpha$: variations less (larger) than $\hat{\sigma}_\alpha$
are considered small (large). This suggests the introduction of dimensionless parameters
\begin{equation}
  \tilde{x}_\alpha
  =
  \frac{x_\alpha}{\hat{\sigma}_\alpha},
\label{eq:dimlessparams}
\end{equation}
which will play a role in the sensitivity analysis of~\sref{sec:Sensitivity}.

The matrix
$(\hat{C}^{-1})_{\alpha\beta}$ can be used to approximate not
only the variances of each parameter but also the correlations between
different parameters. This matrix depends on  the physical
dimensions of the parameters. Rescaling it using the
dimensionless parameters yields the matrix
\begin{equation}
  R_{\alpha\beta} 
  =
  \frac{(\hat{C}^{-1})_{\alpha\beta}}{\sqrt{(\hat{C}^{-1})_{\alpha\alpha}(\hat{C}^{-1})_{\beta\beta}}}
  .
\label{eqn:CorrelationCoefficients}
\end{equation}
The square of the covariances, $R_{\alpha\beta}^2$, defines the coefficients of
determination (CoDs),
which will be discussed in~\sref{sec:CoD}.

\subsection{Calibration strategy and selection of data}
\label{sec:fitdata}

At this point, it is worth recapitulating the history of our  Fayans EDF parameterizations based on careful calibrations of  large, heterogeneous datasets. The
first fit, which was published in \cite{Reinhard2017} and is called $Fy(\Delta
r)$, calibrated the functional without the isovector pairing parameter
$f_{\mathrm{ex},-}^\xi$ and treated pairing at the BCS level.  While the present
study's dataset is an evolution of the datasets from this first fit and from 
\cite{Bollapragada2021}, they are all nearly identical. The BCS pairing
inhibits application of $Fy(\Delta r)$ for weakly bound nuclei. The
next stage aimed to include the  measured charge radius of the
neutron-deficient $^{36}$Ca, which required the use of HFB pairing. The refit
including $^{36}$Ca delivered the parameterization $Fy(\Delta r,{\rm HFB})$ \cite{Miller2019}, which can be applied
without constraints on the binding strength.  Both parameter sets  deliver a fairly good
reproduction of nuclear bulk properties over the chart of nuclei
together with differential charge radii along the   Ca isotopic chain. However,  subsequent
applications revealed that the reproduction of differential radii in heavier
nuclei was deficient. To allow more flexibility, one
must allow different pairing strengths for protons and neutrons, which
amounts to activating the parameter $f_{\mathrm{ex},-}^\xi$. For first
explorations, isotopic radius differences in Sn and Pb were added to the
optimization dataset, which resulted in  a substantial improvement for all isotopic radius
differences without loss in other observables \cite{Karthein2023}.
Here, we  scrutinize the impact of
$f_{\mathrm{ex},-}^\xi$ as such (i.e., without changing the dataset).

In this study we compare the optimization, nonlinear regression analysis, and
sensitivity analysis results for a baseline problem with the Fayans EDF using
$\nx=13$ free model parameters and fixed $\delfxi = 0$ against the $\nx=14$ version
of the baseline problem with $\delfxi$ freed.  The two problems are constructed
with the same dataset, $\dataset$, which  comprises $\nd = 194$
observables that are associated with 69 different spherical, ground-state,
even-even nucleus configurations.  Table~\ref{tab:observableClasses}
shows 
 a breakdown of the  physical observables by class. The energy staggering (last two rows) is defined by means of the three-point energy difference between neighboring even-even isotopes for $\Delta^{(3)}E_n$ and isotones  for $\Delta^{(3)}E_p$. It provides experimental  data to inform the pairing functional. 
The dataset used for optimizing the Fayans EDF consists of
binding energies and their differences and key properties of the charge form factor
\cite{Fri82a} such as charge radius, diffraction (or box-equivalent) radius, and surface
thickness.
The individual data are listed in tables~\ref{tab:fitdata1} and \ref{tab:fitdata2}. 

\begin{table}[htb]
\caption{\label{tab:observableClasses} The classes of physical observables $d_i$ ($i=1,\dots 194$) 
included in this study.}
\begin{center}
{\footnotesize
\lineup
\begin{tabular}{lcc}
\hline
\hline
Class & Symbol & Number of observables\\
\hline
Binding energy & $E_{\rm B}$ & 60\\
Diffraction radius & $R_{\rm box}$ & 28\\
Surface thickness & $\sigma$ & 26\\
Charge radius & $r_{\rm ch}$ & 54\\
Proton single-level energy & $\epsilon_{\mathrm{ls,p}}$ & 3\\
Neutron single-level energy & $\epsilon_{\mathrm{ls,n}}$ & 4\\
Differential radii & $\delta\langle r^2 \rangle$ & 3\\
Neutron radius staggering & $\Delta^{(3)}E_n$ & 5\\
Proton radius staggering & $\Delta^{(3)}E_p$ & 11\\
\hline
\hline
\end{tabular}
%
%
%
%
%
%

}
\end{center}
\end{table}

The adopted errors ($w_i$) associated
with the residuals in the dataset are basically taken from those in~\cite{Reinhard2017}
and are provided in  \ref{app:data}. Their choice is a compromise. 
Typically, the adopted errors are tuned such that the average variance from~(\ref{eq:sigmaa}) fulfills $\hat{\sigma}^2=1$ \cite{Birge1932,DNR14}.
This works only approximately in our case because the model has a systematic error associated with its mean-field approximation neglecting correlations. 
This error has been estimated by computing 
 collective ground-state correlations beyond DFT  throughout the chart of isotopes
 \cite{Klu08a}. The adopted errors are taken from previous fits for which
 the criterion $\hat{\sigma}^2=1$ was approximately fulfilled. 
 The data were selected such that the systematic error remains below the adopted error. The more versatile Fayans functional considered here produces better fits, and one is tempted to reduce the $w_i$ to meet the criterion. But then one may lose
 a great amount of fit data, which would reduce the predictive power of the fit.
 We thus continue to use the inherited adopted errors and accept that we
 deal then typically with $\hat{\sigma}^2\approx 1/4$ for 13D and $\approx 1/5$ for 14D.
A special case is the few data on 
 spin-orbit splittings of single-particle levels; their uncertainty is taken as rather large because 
single-particle energies are indirectly deduced from neighboring odd nuclei, which adds another bunch of uncertainties.

Similar to the previous fits of the Fayans EDF \cite{Reinhard2017,Reinhard2020},
the dataset includes three-point staggering of binding energies for calibrating
pairing properties; see \tref{tab:fitlsgap}.  In this case, however, the dataset includes even-even
staggering as opposed to even-odd staggering; see \cite{Reinhard2017} for more details.

A few crucial differential charge radii 
in Ca
are included in the fit data; see \tref{tab:new-iso}. These were
decisive for determining the advanced gradient terms in the Fayans
EDF related to the parameters $h_{\nabla}^\mathrm{s}$
and $h_{\nabla}^{\xi}$.  For a detailed discussion of the physics
implications see \cite{Reinhard2017}.
The additional data points on differential charge radii
were given small adopted errors to force good agreement for these new data points.

\subsection{Parameter scaling and parameter boundaries}

\begin{table}[!htb]
\lineup
\centering
\caption{The length scales $\sigma_\mathrm{scale}$ used to define the linear scaling of each parameter
at the starting point $\xb_1$, which was the best result
in~\cite{Bollapragada2021}; the result of the 13D optimization
$\hat{\xb}_{13\mathrm{D}}$; and the result of the 14D optimization
$\hat{\xb}_{14\mathrm{D}}$.  The units for
$\rho_\mathrm{eq}$ are in fm$^{-3}$; for $E_{\rm B}/A, K, J,$ and $L$, the units are
in MeV.  All other parameters are dimensionless.}
\label{tab:scaling}

{\footnotesize
\begin{tabular}{cccc}
\br
& $\sigma_\mathrm{scale}(\xb_1)$ & $\sigma_\mathrm{scale}(\hat{\xb}_{13D})$ & $\sigma_\mathrm{scale}(\hat{\xb}_{14D})$ \\
\mr
$\kappa$                & 0.0152  & 0.0121  & 0.0337 \\
$\kappa'$               & 0.546   & 0.606   & 0.452 \\
$a_+^\mathrm{s}$        & 0.0392  & 0.0328  & 0.0438 \\
$h_{\nabla}^\mathrm{s}$ & 0.162   & 0.125   & 0.250 \\
$E_{\rm B}/A$                   & 0.125   & 0.125   & 0.125 \\
$\rho_\mathrm{eq}$      & 0.00484 & 0.00403 & 0.00536 \\
$K$                     & 16.5    & 15.3    & 21.1 \\
$J$                     & 3.18    & 3.04    & 2.70 \\
$L$                     & 33.5    & 29.0    & 19.6 \\
$h_{2-}^\mathrm{v}$     & 27.2    & 78.7    & 10900 \\
$f_{\mathrm{ex},+}^\xi$ & 0.0592  & 0.0450  & 0.0917 \\
$h_{1+}^\xi$            & 0.0832  & 0.0635  & 0.128 \\
$h_\nabla^\xi$          & 0.368   & 0.286   & 0.742 \\
$f_{\mathrm{ex},-}^\xi$ & 2.17    & 2.18    & 0.853 \\
\br
\end{tabular}
}
\end{table}

The model parameters used in the functional have different physical units.  In
addition, empirical studies of the objective function at the different starting
points used in the study reveal that the characteristic length scales of the
objective function along different parameters can vary by several orders of
magnitude at each point and that these length scales can differ
significantly between points.  To aid the optimization and subsequent analysis,
we determined independently at each of several key parameter-space points a linear
scaling of the parameter space such that objective function values change by a
similar amount in magnitude due to offsets along each parameter by the same
amount in the scaled space.  The length scales used to scale our parameter
space are given in \tref{tab:scaling}.

While in a previous \fandf study~\cite{Bollapragada2021}
special techniques were used to maintain optimizations within a constrained region
in which the software was expected to be numerically stable, for this study we
performed only unconstrained optimizations in accord  with (\ref{eqn:min_problem}) and without major issues.

\section{Results}

\subsection{Optimization with \pounders}
The $p=13$ optimization, referred to as 13D in the following, was started from the best result, $\xb_1$, reported
in~\cite{Bollapragada2021}.   The least-squares approximation obtained, called $\xbhata$ (see \tref{tab:PointEstimates}), is different from $\xb_1$ due to improvements made to the software and the aforementioned changes to the dataset. 
The resulting Fayans EDF parameterization is called Fy($\Delta r$, 13D).
\ecnoise tools based on~\cite{more2011ecn,more2011edn,more2014nd} were used to obtain forward-difference approximations to the gradient of the objective function and the Jacobian of the residual function, $\hat{J}(\xbhata) \equiv \Jrhata \in \R^{\nd \times \nx}$, which are needed for assessing the quality of the \pounders solution, nonlinear regression analysis, and sensitivity analysis. 

The $p=14$ optimization, referred to as 14D in the following,  was started from both $\xb_1$ and $\xbhata$, with both
effectively yielding the same least-squares approximation $\xbhatb$; see \tref{tab:PointEstimates}.  For the optimization started at the former
point, the length scale along $\htwovm$ changed significantly enough over the
optimization that the objective function was eventually evaluated at points
where the software failed.  This necessitated determining a new linear scaling
law at an intermediate point and restarting the optimization from that point
using the new scaling.  The gradient and Jacobian, $\Jrhatb$, were obtained with \ecnoise in an identical way to that for the 13D solution.
The resulting Fayans EDF parameterization is called Fy($\Delta r$, 14D).

\begin{figure*}[htbp]
\includegraphics[width=\linewidth]{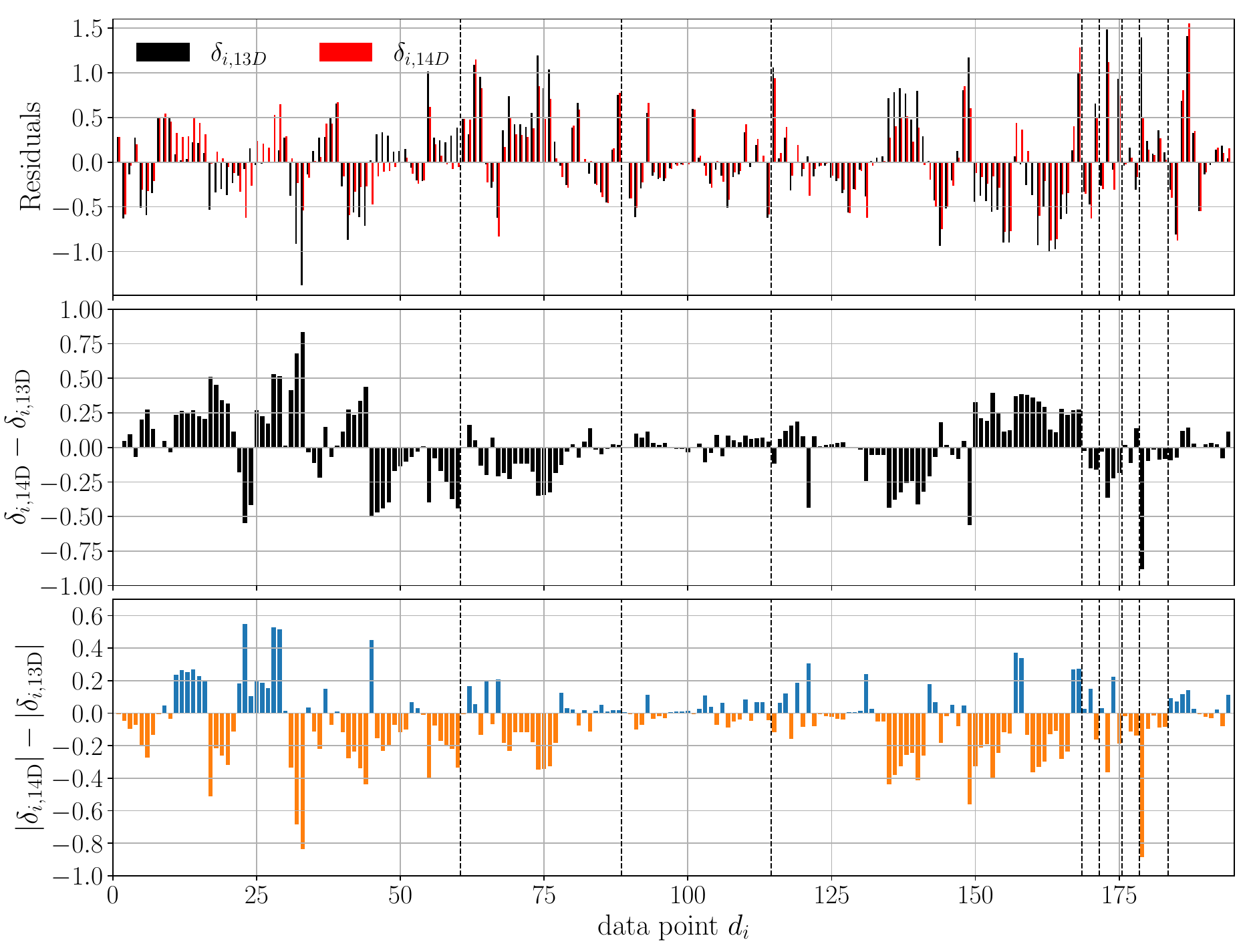}
\caption{(top) Residual  values for the 13D and 14D solutions.
(middle) Change in residual  values between the 13D and 14D solutions.
(bottom) Change in residual   magnitude between
the 13D and 14D solutions.  A negative value indicates that the
magnitude of the associated residual decreased as a result of freeing
$\delfxi$.  The elements are grouped in observable classes of \tref{tab:observableClasses} with an ordering,
from left to right, of $E_{\rm B}$, $R_{\mathrm{box}}$, $\sigma$,
$r_{\mathrm{ch}}$, $\epsilon_{\mathrm{ls,p}}$, $\epsilon_{\mathrm{ls,n}}$,
$\delta\langle r^2\rangle$, $\Delta^{(3)} E_n$, and $\Delta^{(3)} E_p$.
\label{fig:residual_changes}}
\end{figure*}
The top panel of \fref{fig:residual_changes} shows the residuals
elementwise for both solutions. The bottom panel presents the change of
the absolute value of the residuals, with negative (positive) values indicating a gain (loss) in
quality of the agreement to data. The residuals are grouped by classes of observables with
subgrouping into isotopic or isotonic chains where possible. Large changes
between the two parameterization are seen for binding energies and charge radii, moderate
changes for diffraction radii, and small differences for surface thicknesses. The
bottom panel shows that the extension from 13D to 14D, while generally beneficial,
can decrease agreement with experiment
for some observables. To quantify this effect, we now inspect partial sums of the objective function $f$ rather than single
residuals.

\begin{figure}
\centering
\includegraphics[width=0.7\linewidth]{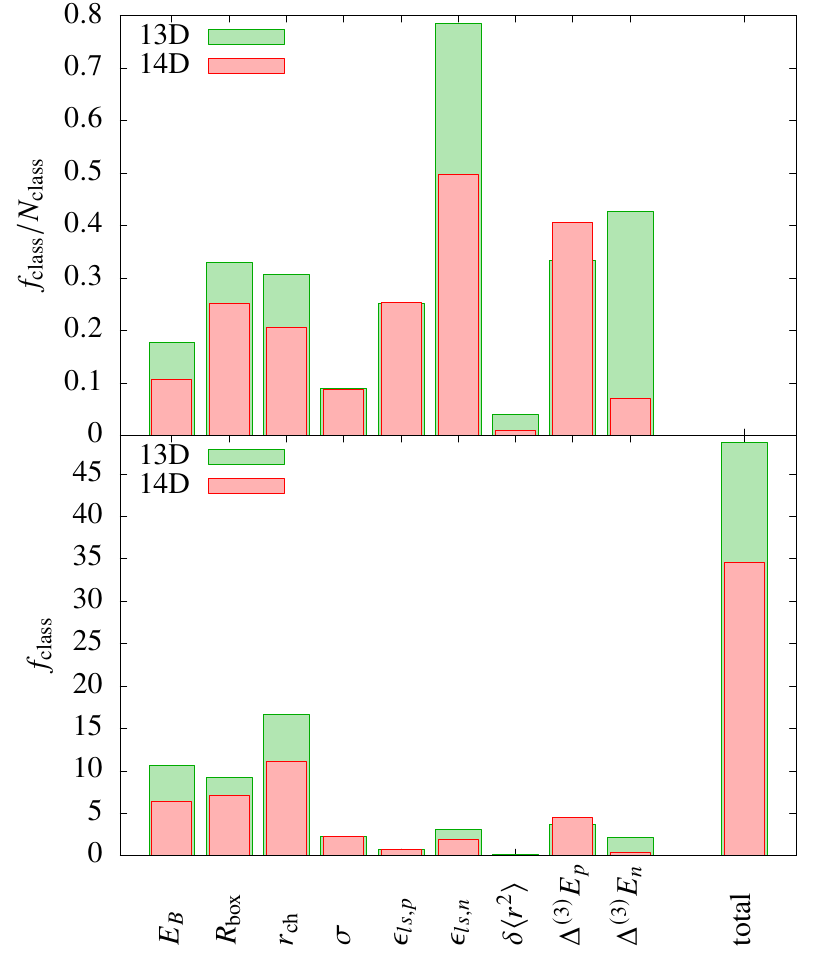}
\caption{\label{fig:Chi2Comparison} Breakdown of the contributions to
  the total objective function $f$ by observable class (see \tref{tab:observableClasses}) for the Fy($\Delta r$,13D) and Fy($\Delta r$,14D)
  parameterizations (see \tref{tab:PointEstimates}). Presented are
  the (bottom) summed contribution $f_\mathrm{class}$ within a class and the (top)
  average contribution per data point $f_\mathrm{class}/N_\mathrm{class}$,
  where $N_\mathrm{class}$ is the number of data points in the given class.
}
\end{figure}
The lower panel of \fref{fig:Chi2Comparison} shows the total objective function $f$ (rightmost bar) and the partial contributions $f_\mathrm{class}$ summed over each class of observables (energy, radii, etc.) as indicated. The upper panel complements the information by showing the $f_\mathrm{class}$ per data point for each class. 
Adding $f_{\mathrm{ex},-}^\xi$ to the set of optimized parameters results in
a clear gain in quality for
most observables.
Several observables (surface thickness, proton spin-orbit splitting, and proton gap)
are hardly affected by this change. The most significant improvement is seen for the neutron gap.

The $\chi^2$ per datum (upper panel) shows that the optimization resulted in values considerably below one. This is due to our choosing to take the correlation effects as a guideline for the adopted errors.
All in all, the total $\chi^2$ has been reduced by about 30\% through the introduction of $\delfxi$. This  surprisingly large gain  suggests that the new feature brought in, namely, to allow different pairing strengths for protons and neutrons, is physically significant.

%
\begin{table}[tbp]
\caption{The least-squares estimate of the 13D (top) and 14D (bottom) optimization problems and the standard deviations defined in (\ref{eq:sigmaa}) that partially characterize
the approximated zero-mean normal distributions of the associated parameter estimation error.  $\rho_\mathrm{eq}$ is in fm$^{-3}$;  the units of $E_{\rm B}/A, K, J, L$
are in MeV; other parameters  are dimensionless.}
\lineup
\centering
{\footnotesize
\begin{tabular}{lcc}
\br
 & $\hat{x}_\alpha$(13D) & $\hat{\sigma}_\alpha$\\
\mr
$\kappa$                & \0\00.190867 & 0.002024\\
$\kappa'$               & \0\00.032788 & 0.014017\\
$a_+^\mathrm{s}$        & \0\00.564916 & 0.021191\\
$h_{\nabla}^\mathrm{s}$ & \0\00.408625 & 0.089848\\
$E_{\rm B}/A$                   & \0\-15.873321 & 0.014744\\
$\rho_\mathrm{eq}$      & \0\00.165064 & 0.000763\\
$K$                     & 203.587853 & 7.638661\\
$J$                     & \029.069702 & 0.639137\\
$L$                     & \044.228119 & 6.477113\\
$h_{2-}^\mathrm{v}$     & \015.325767 & 6.456659\\
$f_{\mathrm{ex},+}^\xi$ & \0\0\-3.963726 & 0.175008\\
$h_{1+}^\xi$            & \0\03.540660 & 0.215688\\
$h_\nabla^\xi$          & \0\03.270458 & 0.191246\\
\br
\end{tabular}
}

{\footnotesize
\begin{tabular}{lcc}
\br
 & $\hat{x}_\alpha$(14D) & $\hat{\sigma}_\alpha$\\
\mr
$\kappa$                & \0\00.185929   & \0\00.002038\\
$\kappa'$               & \0\00.019272   & \0\00.014026\\
$a_+^\mathrm{s}$        & \0\00.538812   & \0\00.016033\\
$h_{\nabla}^\mathrm{s}$ & \0\00.307605   & \0\00.072431\\
$E_{\rm B}/A$                   & \0\-15.881322  & \0\00.010785\\
$\rho_\mathrm{eq}$      & \0\00.164331   & \0\00.000648\\
$K$                     & 214.169984     & \0\06.062988\\
$J$                     & \030.248343    & \0\00.432775\\
$L$                     & \062.427904    & \0\03.181482\\
$h_{2-}^\mathrm{v}$     & 406.608365     & 486.788920\\
$f_{\mathrm{ex},+}^\xi$ & \0\0\-4.315720 & \0\00.169836\\
$h_{1+}^\xi$            & \0\03.983162   & \0\00.205909\\
$h_\nabla^\xi$          & \0\03.532572   & \0\00.281308\\
$f_{\mathrm{ex},-}^\xi$ & \0\0\-0.357833 & \0\00.063162\\
\br
\end{tabular}
}
\label{tab:PointEstimates}
\end{table}
\Tref{tab:PointEstimates} shows the model parameters 
of Fy($\Delta r$, 13D) and Fy($\Delta r$, 14D)
together with their approximated standard deviations. The differences of the parameter values between the two calibrations stay more
or less within these standard deviations. An exception is the parameter $f_{\mathrm{ex},-}^\xi$, which is specific to 14D. Its value is much larger than its standard deviation, meaning that it is not compatible 
with 13D parameterizations that set $f_{\mathrm{ex},-}^\xi=0$. The model parameters for the volume terms are expressed by NMP. Their actual values agree nicely with the commonly accepted values; see the discussion in \cite{UNEDF2}. The largest difference between 13D and 14D
is seen in the value of 
$h_{2-}^\mathrm{v}$, which is
already large for 13D and grows much larger for 14D. But one should not be misled by the dramatic change in value. A large $h_{2-}^\mathrm{v}$ simply renders the  second term in the denominator of the isovector volume term in (\ref{EFay-dens}) all-dominant such that large changes have only small effect.
This parameter is extremely weak in the regime of large values. As a consequence, its computed variance is large and exceeds the bounds of the linear regime. One should not take this variance literally; it is simply a signal of a weakness of the model in this respect.

The  strengths of the density-independent pairing functional
$f_{\mathrm{ex},\pm}^\xi$ define the
density-independent proton pairing strength
$f^\xi_{\mathrm{ex},p}=f_{\mathrm{ex},+}^\xi+f_{\mathrm{ex},-}^\xi$
and  density-independent neutron pairing strength
$f^\xi_{\mathrm{ex},n}=f_{\mathrm{ex},+}^\xi-f_{\mathrm{ex},-}^\xi$.
According to \tref{tab:PointEstimates},  this
yields $f^\xi_{\mathrm{ex},p}=f^\xi_{\mathrm{ex},n}=-3.963726$ for 13D
and  $f^\xi_{\mathrm{ex},p}=-4.673553,f^\xi_{\mathrm{ex},n}=-3.957887$ 
for 14D. This means that the  density-independent neutron pairing strength remains practically unchanged when going from 13D to 14D  while the magnitude of the proton
strength significantly increases. 
This result is typical for all modern Skyrme functionals 
\cite{Bertsch2009,UNEDF2}. It is 
satisfying that the Fayans functional behaves the same way. 

\subsection{Correlations between observables/parameters}
\label{sec:CoD}

\begin{figure}[htb]
\centering{
\includegraphics[width=0.65\linewidth]{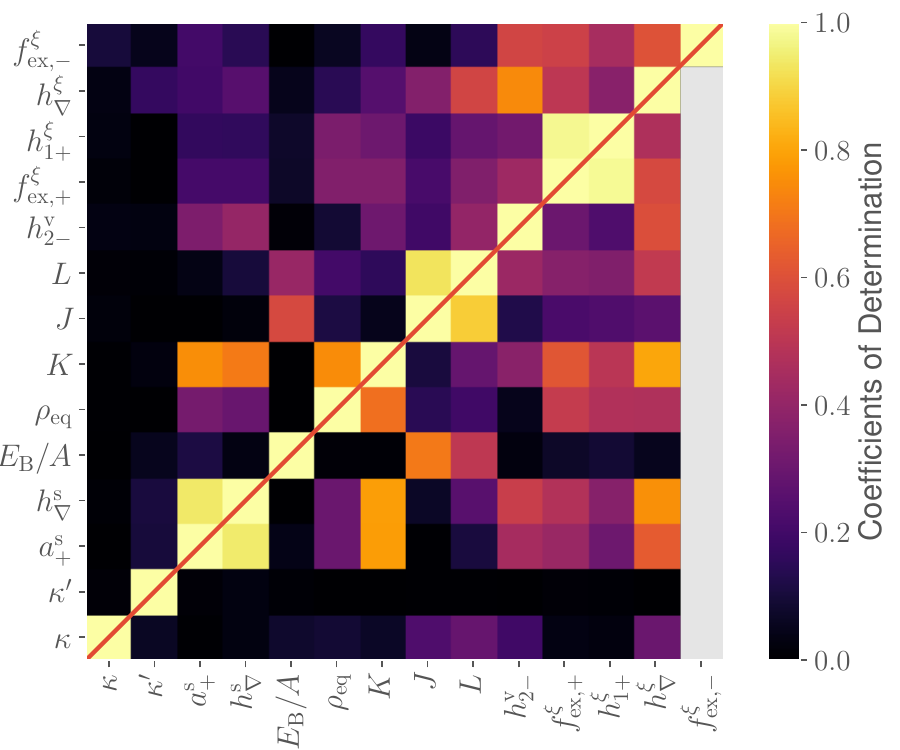}
}
\caption{Coefficients of determination $R_{\alpha \beta}^2$ for the 13D  (lower triangle)
and 14D  (upper triangle) calibrations.
The parameters are ordered to highlight their correlations.
\label{fig:stat_matrices}
}
\end{figure}
The correlations between model parameters in the vicinity of our solution are quantified by the
matrix of CoDs.
\Fref{fig:stat_matrices} visualizes the correlations for both the 13D and 14D calibrations.  
Considerable correlations exist for some groups of parameters, which
show  that the number of the degrees of
freedom of the model is less than the  number of
parameters~\cite{Kejzlar_2020}.
For example, strong correlations
exist between the two surface parameters 
($a^\mathrm{s}_+,h^{\mathrm{s}}_\nabla$), between the two symmetry parameters ($J$,$L$), 
and between two pairing parameters ($f^\xi_{\mathrm{ex},+},h^\xi_{1+}$).  
Several somewhat smaller, but still strong, correlations also exist. For example, surface parameters and $K$ 
correlate because both have impact on nuclear radii. 
Binding energy and symmetry energy parameters correlate because of some long isotopic 
chains in the data pool. Practically uncorrelated are the two spin orbit parameters $\kappa$ and $\kappa'$. All these correlations behave similarly in
both calibration variants, and they appear also in other models \cite{Erler_2015}. Not surprisingly, however, some correlations differ with pairing parameters. For example, the 13D variant
shows considerable correlation of $f^\xi_{\mathrm{ex},+},h^\xi_{1+}$ with surface parameters while the 14D variant has lost this correlation because of
the introduction of the isovector pairing parameter $f^\xi_{\mathrm{ex},-}$. A similar reduction of correlations happens for the connection between pairing parameters and the group $K$, $\rho_\mathrm{eq}$. It
is not uncommon for correlations to get 
reduced with new parameters because they remove a previously existing rigidity within a model \cite{Reinhard2010s,Rei22d}. Although the new
parameter $f_{\mathrm{ex},-}^\xi$ has most of its correlations within the group of pairing parameters, it
is rather independent from them. Correlations with other model parameters are
generally weak, except for 
$h^v_{2-}$, which is related to isovector density dependence.

\subsection{Sensitivity analysis}
\label{sec:Sensitivity}

Minimization of the objective function delivers the optimized
parameter set $\xbhat$. Sensitivity analysis deals with the question
of how the parameters change, 
$\tilde{x}_\alpha\longrightarrow \tilde{x}_\alpha+\delta\tilde{x}_\alpha$, 
if the data are varied by a small amount,
$\tilde{d}_i\longrightarrow\tilde{d}_i+\delta\tilde{d}_i$. 
Note that we formulate the problem in terms of dimensionless data (\ref{eq:dimlessdata}) and dimensionless parameters (\ref{eq:dimlessparams}) 
to allow a seamless combined handling of different types of data and parameters.
Following
forward error analysis~\cite{Bjorck96}, we search for the solution
$\hat{\xb} + \delta \xb$ to the optimization problem 
(\ref{eqn:min_problem}) but with
the modified dataset  $\tilde{d}_i + \delta\tilde{d}_i$ and find
\begin{equation}
  \delta\tilde{x}_\alpha^{(i)}
  =
  S_{\alpha i}
  \delta\tilde{d}_i
  \;,\;
  S_{\alpha i}
  =
  \frac{\left[(\hat{J}^T \hat{J})^{-1}\hat{J}^T\right]_{\alpha i}}{\hat{\sigma}_\alpha}. 
\label{eq:senitivity_define}
\end{equation}
\Eref{eq:senitivity_define} establishes the connection to a parameter
change for small perturbations $\delta\tilde{d}_i$, and can be expressed also as
$S_{\alpha i}=\delta\tilde{x}_\alpha^{(i)}/\delta\tilde{d}_i$. In the following, we assume that all dimensionless data points are changed by the same amount
$\delta\tilde{d}_i=\delta\tilde{d}=$ constant. Since (\ref{eq:senitivity_define}) is in the linear regime, changes $\delta\tilde{x}_\alpha^{(i)}$ are proportional to $\delta\tilde{d}$. We are interested in the relative effects, and thus the actual value of $\delta\tilde{d}$ is unimportant once the approximation in (\ref{eq:senitivity_define}) is employed.

From the (dimensionless)
sensitivity matrix we build the real-valued, positive number
$\mathcal{S}_{\alpha i}=|S_{\alpha i}|^2$ as a measure for the impact
of data point $\tilde{d}_i$ on parameter $\tilde{x}_\alpha$.
The matrix of sensitivities $\mathcal{S}_{\alpha i}$ carries a huge amount
of information about the calibrated model.
First, we look at the sensitivity for observable classes $C$ of energies,
radii, and so on. Instead of asking, for example, what is the impact of the energy
of $^{208}$Pb on a parameter $\tilde{x}_\alpha$, we ask now, what is
the impact of all the energy entries. To that end, we build the sum
of the detailed $\mathcal{S}_{\alpha i}$ over the data $i$ in class
$C$:
\begin{equation}
     \mathcal{S}_\alpha^{(C)}=\sum_{i\in C}
     \mathcal{S}_{\alpha i}.
\end{equation}
The relative sensitivity per class is given by
\begin{equation}
  s_\alpha^{(C)}
  =
  \frac{\mathcal{S}_\alpha^{(C)}}{\sum_c\mathcal{S}_\alpha^{(c)}}
\label{eq:rel_sens_class}
\end{equation}
and does not depend on the choice of $\delta\tilde{d}$ as desired.
\begin{figure}[tbh]
\centering
\includegraphics[width=0.75\linewidth]{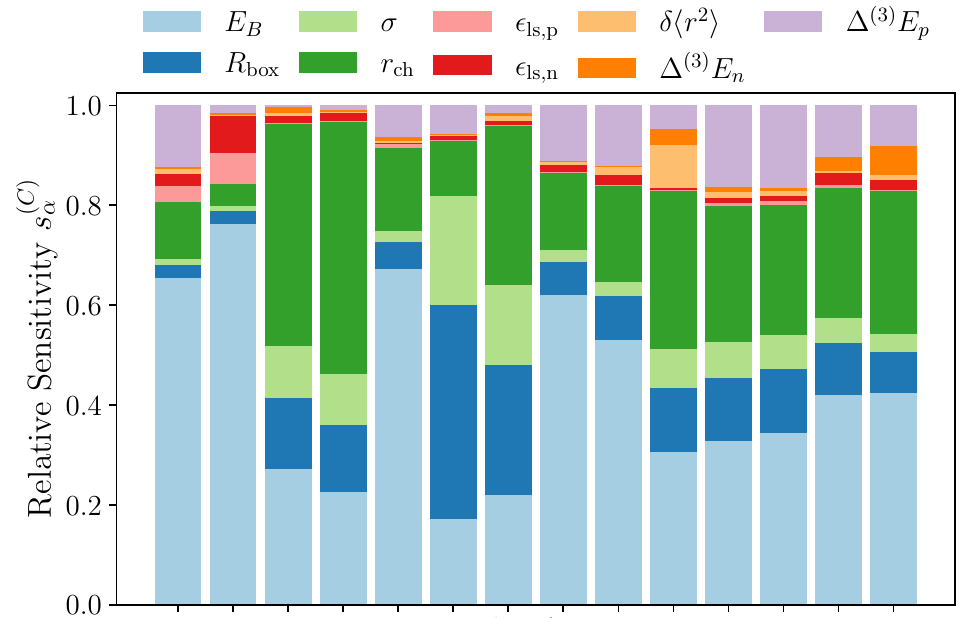}\\
\includegraphics[width=0.75\linewidth]{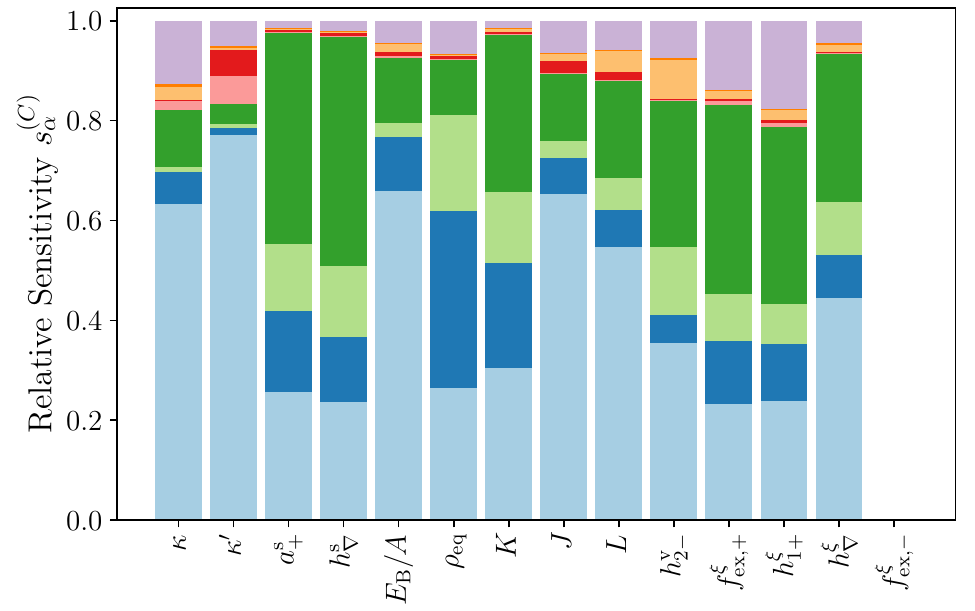}
\caption{Relative sensitivities per data class (\ref{eq:rel_sens_class})
for the model parameters of the 13D Fy($\Delta r$,13D) (bottom) and 14D Fy($\Delta r$,14D)
(top) EDFs. The data classes are represented by colors as indicated.
The $\delta\langle r^2\rangle$ represent the isotopic radius$^2$ differences and the
$\Delta^{(3)}E$ the odd-even staggerings of energies.
\label{fig:sensitivity_by_class}
}
\end{figure}
\Fref{fig:sensitivity_by_class} shows the relative sensitivities for the 13D
and 14D calibration variants. The patterns are similar to those already seen
for Skyrme models \cite{UNEDF0}.  The parameters $E_{\rm B}/A$, $J$, and $L$ are most influenced by the
binding energy data while 
$\rho_\mathrm{eq}$, $K$, and surface parameters
$a_+^s$ and $h_\nabla^s$ are more sensitive to  surface data $R_{\rm box}, \sigma$, and $r_{\rm ch}$. The
spin orbit parameters $\kappa$ and $\kappa'$ are dominated by energy
information while the  data on the 
spin-orbit
splitting, $\epsilon_\mathrm{ls}$, play a surprisingly
small role. The
pairing parameters 
$f^\xi_{\mathrm{ex},\pm},h^\xi_{1+},$ and 
$h^\xi_\nabla$ are  impacted primarily by binding energies and surface data.
 The differential data, $\delta\langle r^2\rangle$ and $\Delta^{(3)}E$, are important for the determination of the pairing functional in the 14D variant, especially for $h^\xi_{1+}$.

The effect of one data point
$\tilde{d}_i$ on the model parameters also provides interesting information. To this end, we add up the
detailed sensitivities over all parameters, coming to the total impact
of a data point $\tilde{d}_i$ as $\sum_\alpha\mathcal{S}_{\alpha i}$.
To render the different data points comparable, we use a constant
change $\delta\tilde{d}_i=\delta\tilde{d}=1$. To see the effect of another value
$\delta\tilde{d}$, we simply scale the resulting total impact by this value.

\begin{figure}[htbp]
\centering
\includegraphics[width=1.0\linewidth]{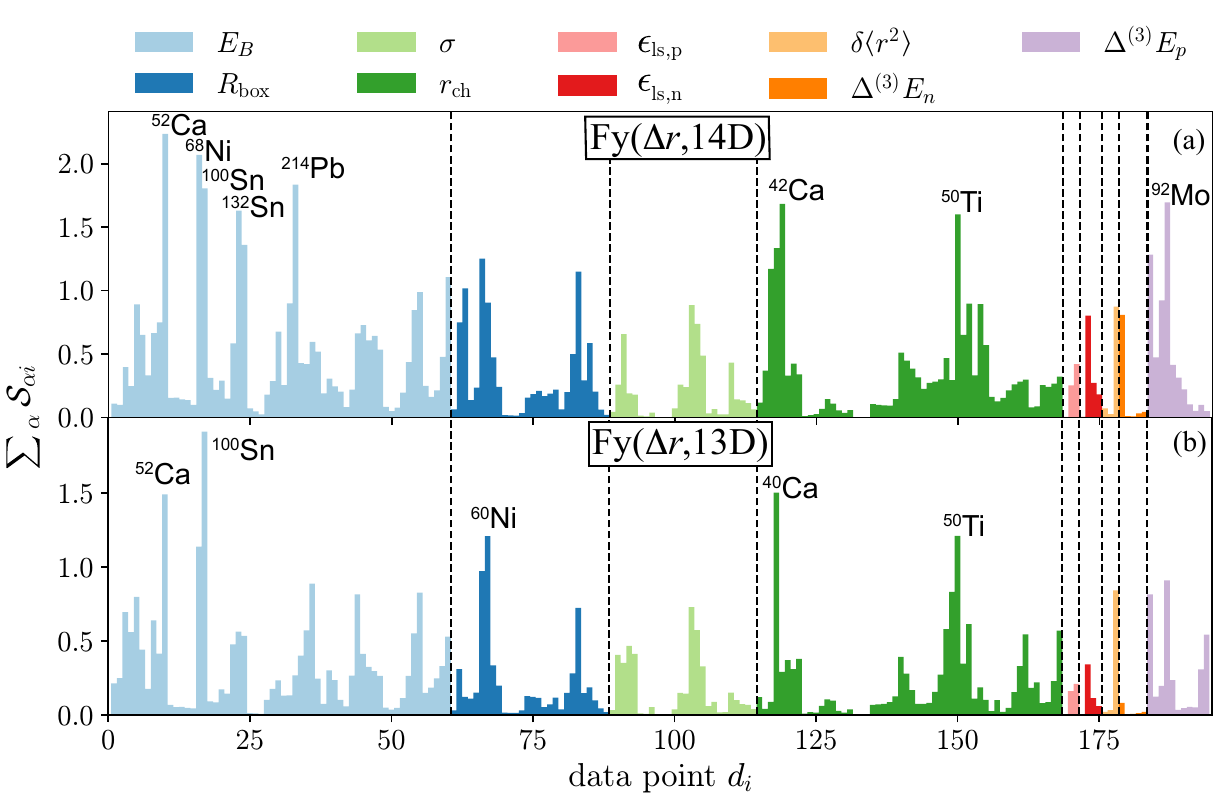}
\caption{Total impact of a
  data point $i$ on the  parameters 
  of the Fy($\Delta r$,14D)
  (a) and Fy($\Delta r$,13D) (b) EDFs.
The data classes are separated by dashed  vertical lines as in \fref{fig:residual_changes}. The data points having the largest impact on calibration results are indicated.
}
\label{fig:sensitivity_by_observable13}
\end{figure}

\Fref{fig:sensitivity_by_observable13} shows the result of our sensitivity study. Note that the absolute values are unimportant here; the main information is contained in
the relative distribution. In general, the calibration dataset is fairly balanced, with only several  data points showing significant variations.
The most pronounced peaks in the 14D variant are  the binding energies
of $^{52}$Ca, $^{68}$Ni, $^{100}$Sn, and $^{214}$Pb; the charge radii
of $^{42}$Ca and $^{50}$Ti; and the proton 3-point  binding energy difference for $^{92}$Mo.
For the 13D EDF,  the importance of $E_{\rm B}$ for $^{132}$Sn and
$^{214}$Pb
and $\Delta^{(3)}E_p$ for $^{92}$Mo is reduced.
Furthermore, we note that the sensitivities for $\Delta^{(3)}E_n$, and even more so for $\Delta^{(3)}E_p$, are generally larger for 14D. These results are related to the fact that 14D has more leeway in the pairing functional. The results show, first, that sensitivity  not only is a property of data but also is intimately connected with the form of the functional and, second, that more versatility in the functional often leads to more sensitivity.

\section{Predictions}

\subsection{Impact of isovector pairing on pairing gaps}
\label{sec:gaps}

At the end of the discussion of \tref{tab:PointEstimates}, we
saw that the density-independent proton pairing strength is increased when going
from 13D to 14D while the neutron strength remains almost the same.
This should be visible from typical calculated pairing observables (e.g., 
the proton and neutron
pairing gaps).

\begin{figure}[htb]
\centering
\includegraphics[width=0.6\linewidth]{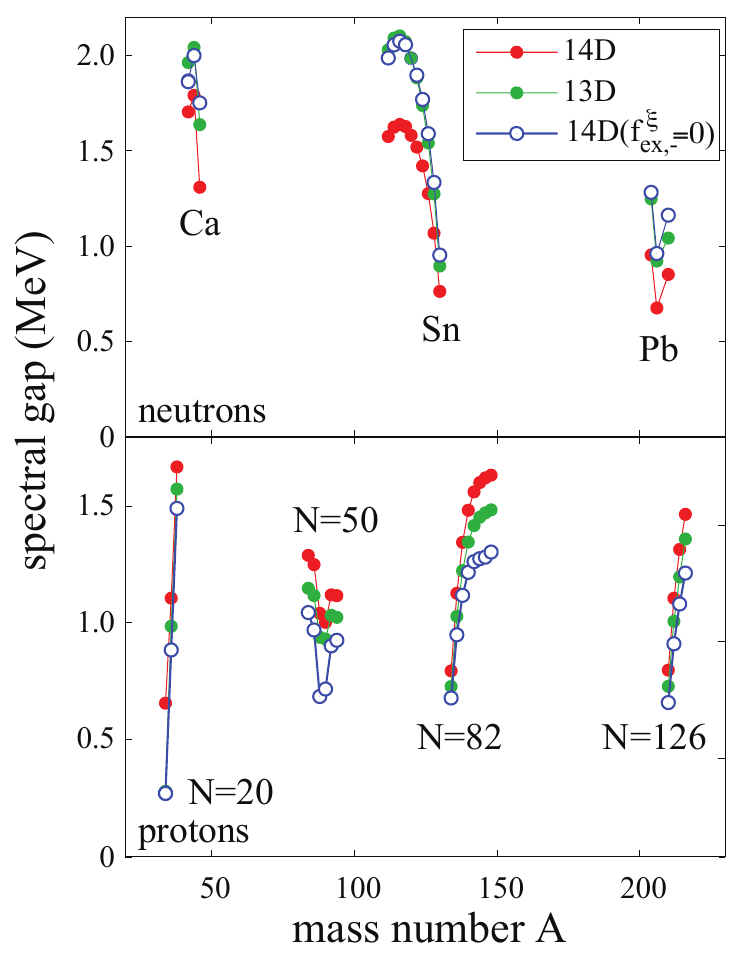}
\caption{ Comparison of the spectral pairing gaps \cite{Ben00a}, 
$\overline{\Delta}_{\tau_3}=
\sum_{\alpha\in {\tau_3}}
\Delta_\alpha u_\alpha v_\alpha /\sum_{\alpha\in {\tau_3}}
u_\alpha v_\alpha$, for (top) neutrons and (bottom) protons
obtained with
 Fy($\Delta r$,14D) and
Fy($\Delta r$,13D) and also with  Fy($\Delta r$,14D) with $f_{\mathrm{ex},-}^\xi=0$. 
\label{fig:compare-gaps-3}}
\end{figure}

\Fref{fig:compare-gaps-3} compares the 
spectral pairing gaps \cite{Ben00a}
obtained with
 Fy($\Delta r$,14D) and
Fy($\Delta r$,13D) and  also with Fy($\Delta r$,14D) assuming  $f_{\mathrm{ex},-}^\xi=0$. As expected, when going from 13D to 14D, proton gaps increase. However, the
neutron gaps decrease substantially from 13D to 14D while the density-independent pairing strengths are practically the same in both variants. This result indicates that 
the rearrangement of all parameters,
in particular  those defining the  density-dependent part of the pairing functional, strongly impact 
 spectral pairing gaps.
As a counter check, we also considered a variation of
14D with the only change that we fix
$f^\xi_{\mathrm{ex},-}=0$.  The difference between the results of the
13D  variant and those of the 14D variant having $f^\xi_{\mathrm{ex},-}=0$ indicates the impact of  readjustment of 13 parameters of 13D in the 14D results.

\subsection{Predictions of observables along isotopic chains}
\label{sec:chains}

As discussed earlier, the additional isovector  degree of
freedom in Fy($\Delta r$,14D) allows a  better adjustment to data, particularly with regard to 
isovector trends.
This raises the question of how the two 
parameterizations perform in
extrapolations outside the pool of the training dataset
 $\cal D$.
We look at this now
in terms of four long isotopic chains of spherical semi-magic nuclei: Ca, Sn, and Pb. We also study the  deformed chain of Yb isotopes. 

\begin{figure}[htb]
\centering
\includegraphics[width=\linewidth]{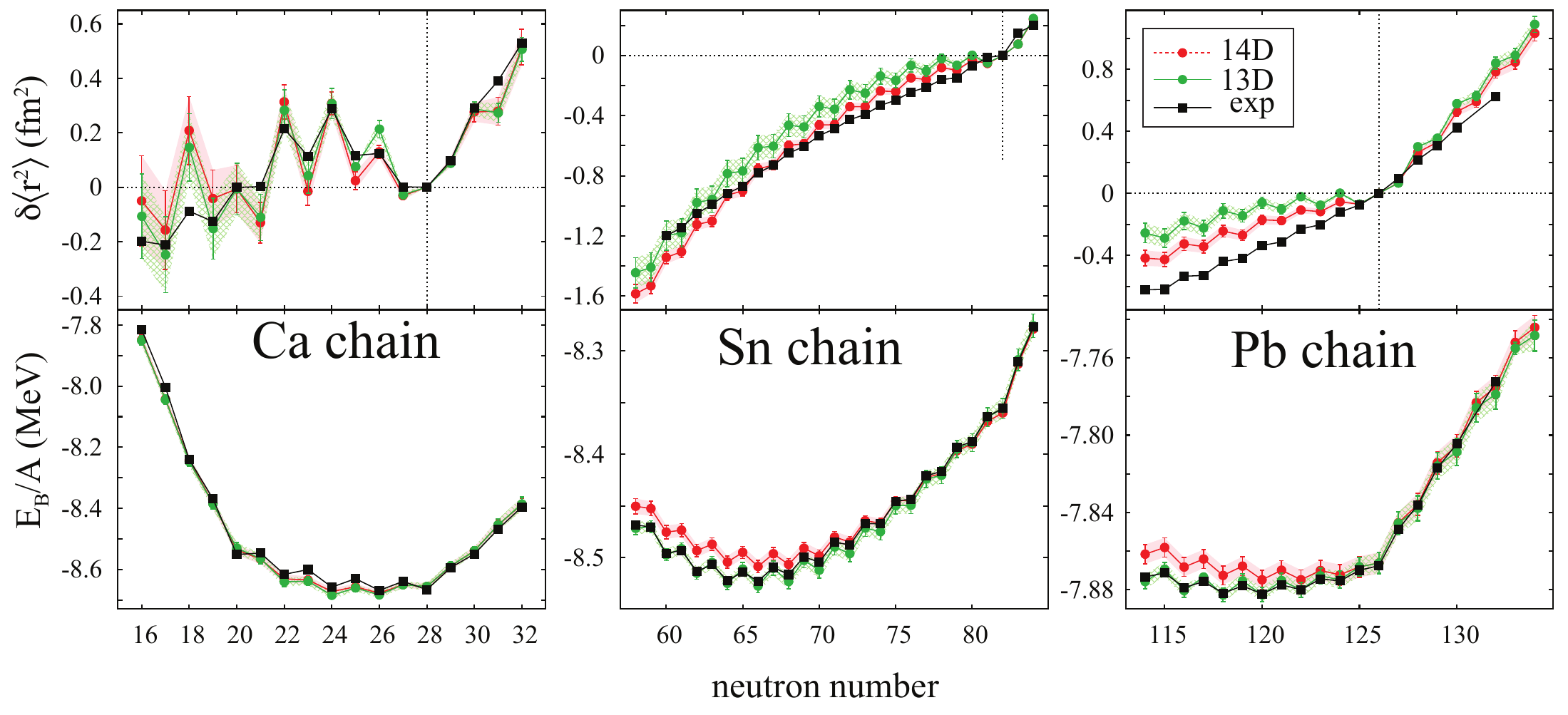}
\caption{ Comparison of the Fy($\Delta r$,13D) and Fy($\Delta r$,14D)  results 
for  $\langle\delta r^2\rangle$ (top) and $E_{\rm B}/A$ (bottom) 
with experiment for three semi-magic isotopic chains: Ca (left), Sn (middle), and Pb (right).
  The statistical uncertainties of the predictions \cite{DNR14} are shown as error bars and error bands.
  All even-even nuclei in these chains are spherical, and calculations
  were done with the axial  DFT solver {\tt
  SkyAx}. 
The differential radii are shown
relative to $^{48}$Ca, $^{132}$Sn, and $^{208}$Pb. Experimental binding energies are from~\cite{Aud02aR}. Experimental radii are from \cite{Fri82a,GarciaRuiz16} (Ca),  \cite{Yordanov2020} (Sn), and
\cite{Angeli:2013} (Pb).
\label{fig:comp1314-comb}}
\end{figure}
Figure~\ref{fig:comp1314-comb} shows binding energies and differential radii
 along the Ca, Sn, and Pb chains. As expected, binding energies
are  well described for the fit nuclei, which are the even-even isotopes
$^{40}$Ca-$^{48}$Ca, Sn with $N\ge72$, and Pb with $N\ge122$. The
agreement persists along the whole Ca chain. Differences develop at
the lower ends of the Sn and Pb chains where 13D remains close to data
and 14D becomes slightly less bound. This happens because 14D produces
less pairing for the proton-rich isotopes   than does 13D, a consequence of
the isovector pairing.  This
should not be taken too seriously because the low-$N$ isotopes are becoming
increasingly deformation-soft and thus prone to ground-state correlations.

The differential charge radii are shown the upper panels in \fref{fig:comp1314-comb}.  This observable is
more sensitive to isovector properties than the absolute charge radii.
The trends in the Ca chain are similar for 13D and 14D. Both tend to slightly
overestimate the odd-even staggering of 
radii. This
is a feature already known from earlier Fayans EDF studies
\cite{Reinhard2017,Miller2019}.  Note, however, that the odd-even charge radius staggering
had not been included in the dataset $\cal D$.   
The overall trend of differential radii for Sn and Pb is similar to that 
for energies, with an increasing difference between 13D and 14D toward low
$N$. For both chains, the Fy($\Delta r$,14D) results stay closer to data.  We note that the charge-radius kink at $^{208}$Pb is  heavily influenced
by the pairing and surface effects \cite{Rei95f,Gorges2019}.

\begin{figure}[htb]
\centering
\includegraphics[width=0.4\linewidth]{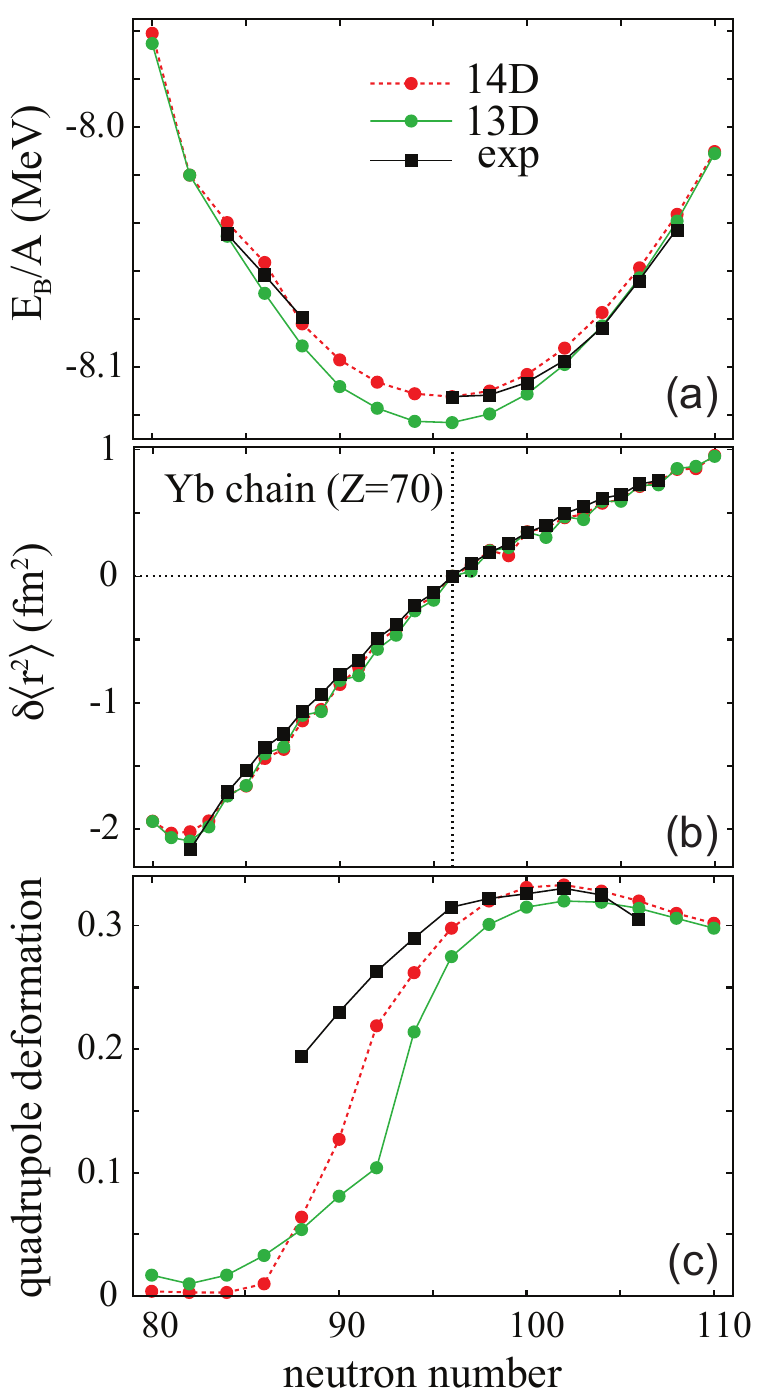}
\caption{ Comparison of the Fy($\Delta r$,13D) and Fy($\Delta r$,14D)  results with experiment (where available) for the chain of Yb
  isotopes ($Z=70$). Since most of these isotopes are deformed,
  calculations were performed with the axial DFT solver {\tt
  SkyAx}. 
  (a) Total binding energy per particle. 
  (b)  Differential radii relative to $^{166}$Yb.
  (c) Proton quadrupole deformations.
  Binding energies and deformations are calculated for
  even-even isotopes only.  
  Experimental values are taken from \cite{Aud02aR} (binding energies), 
  \cite{Angeli:2013} (charge radii), and \cite{Ram01a}
 (deformations). 
\label{fig:comp1314-Yb}}
\end{figure}
Our calibration dataset ${\cal D}$
consists of data on spherical nuclei. It is thus interesting to look at the performance
of Fy($\Delta r$,13D) and Fy($\Delta r$,14D) for well-deformed nuclei.
\Fref{fig:comp1314-Yb} shows binding energies, differential radii,
and proton quadrupole deformations along the chain of Yb isotopes containing many deformed nuclei. 
For deformed systems, we  augment the 
binding
energies by a rotational energy correction approximating  the angular
momentum projection results as outlined in \cite{Rei78a,Erl11aR}. This
correction vanishes for spherical nuclei as discussed earlier.
The calculated binding energies 
agree with the data, especially    near the spherical $^{152}$Yb and for the well-deformed
heavier isotopes. Small differences are seen in the transitional
region. 
As in our previous studies \cite{Reinhard2022q,Hur2022}, the description of differential  radii is excellent.
The proton quadrupole deformations $\beta_{2,p}$ show a transition from spherical shapes
near the semi-magic $^{152}$Yb to well-deformed isotopes for
$N>95$. We note that experimental  $\beta_2$
deformations deduced from $B(E2)$  values include  zero-point  quadrupole fluctuations from ground-state vibrations. The
latter are particularly large in transitional nuclei. A detailed
comparison with data would require  accounting  for these fluctuations.

Summarizing this section, the Fayans functionals  Fy($\Delta r$,13D) and Fy($\Delta r$,14D) calibrated in this work perform well on the testing set of  observables for spherical and deformed nuclei.
In general the 14D model performs slightly better, especially for charge radii.

\section{Conclusions}
In  previous work \cite{Bollapragada2021} we studied
 the performance of optimization-based training algorithms in the context of computationally expensive nuclear physics models based on modest calibration datasets. 
We concluded that the \pounders  algorithm, within a budget of function evaluations, is extremely robust in the context of nuclear EDF calibration.

In this work we employed \pounders 
to
carry out parameter estimation of two Fayans functionals,   Fy($\Delta r$,13D) and Fy($\Delta r$,14D). The latter functional accounts for different strengths of proton and neutron pairing, which generally improved the agreement of the model with ground-state properties.

We carried out sensitivity analysis of these 13D and 14D 
parameterizations and studied 
the sensitivity of model parameters to changes in data points
${d}_i$. We concluded that the binding energy
of $^{52}$Ca,  $^{68}$Ni, $^{100}$Sn, and $^{214}$Pb, the charge radii of
of $^{42}$Ca and $^{50}$Ti, and the proton 3-point  binding energy difference for $^{92}$Mo have the most pronounced impact on Fy($\Delta r$,14D).

In future work we will generalize the surface Fayans functional by adding the isovector surface term. Such an extension is important for systematic
calculations of deformed nuclei and fission \cite{Nikolov2011,UNEDF1}. New calibration datasets will include data on deformed nuclei, including fission isomers.

\section*{Acknowledgments}
This material was based upon work 
supported by the U.S.\ Department of
Energy, Office of Science, Office of Advanced Scientific Computing
Research, applied mathematics and SciDAC NUCLEI programs under Contract Nos.\
DE-AC02-06CH11357 and DE-AC02-05CH11231.  This work was also supported by the U.S.\ Department of Energy, Office of Science, Office of Nuclear Physics under award numbers DE-SC0013365 and DE-SC0023688 (Michigan State University), and DE-SC0023175 (NUCLEI SciDAC-5 collaboration).
We gratefully acknowledge the computing resources provided on Bebop, a 
high-performance computing cluster operated by the Laboratory Computing Resource 
Center at Argonne National Laboratory.

\clearpage

\appendix

\section{Local densities and currents in detail}
\label{sec:dens}

The Fayans EDF, as the Skyrme EDF, is formulated in
terms of  local densities and currents. 
\begin{equation}
  \begin{array}{rclcl}
  \hline
  \mbox{Symbol} && \mbox{Expression} 
  && \mbox{Name} 
  \\[5pt]
   \rho_{\tau_3}& = &\displaystyle
     \sum_{\alpha\in {\tau_3}} v_{\alpha}^2|\varphi_{\alpha}|^2
  &\;\;&
  \mbox{density}
\\
   \vec{s}_{\tau_3}& = &\displaystyle
     \sum_{\alpha\in {\tau_3}} v_{\alpha}^2
       \varphi^+_{\alpha} \hat{\vec{\sigma}}\varphi^{\mbox{}}_{\alpha}
  &\;\;&
  \mbox{spin density}
  \\
   \vec{j}_{\tau_3}& = &\displaystyle
   \Im{m}\left\{\sum_{\alpha\in {\tau_3}} v_{\alpha}^2 \varphi^+_{\alpha}
              \vec{\nabla}\varphi^{\mbox{}}_{\alpha}\right\}
  &\;\;&
  \mbox{current}
  \\
   \vec{J}_{\tau_3} & = &\displaystyle
    -\mathrm{i}\sum_{\alpha\in {\tau_3}} v_{\alpha}^2
             \varphi_{\alpha}^+
                \vec{\nabla} \times \hat{\vec{\sigma}} \varphi^{\mbox{}}_{\alpha}
  &\;\;&
  \mbox{spin-orbit density}
\\
   \tau_{\tau_3}& = &\displaystyle
     \sum_{\alpha\in {\tau_3}} v_{\alpha}^2|\vec{\nabla}\varphi_{\alpha}|^2
  &\;\;&
  \mbox{kinetic-energy density}
\\
   \vec{\tau}_{\tau_3} & = &\displaystyle
    -\mathrm{i}\sum_{\alpha\in {\tau_3}} v_{\alpha}^2
            \vec{\nabla}\varphi_{\alpha}^+
           \cdot\vec{\nabla} \,\hat{\vec{\sigma}} \varphi^{\mbox{}}_{\alpha}
  &\;\;&
  \mbox{kinetic spin-density}
  \\[12pt]
   \xi_{\tau_3} 
   &=& \displaystyle
     \sum_{\alpha\in {\tau_3},\alpha>0}f_\alpha  u_{\alpha}v_{\alpha}|\varphi_{\alpha}|^2
  &\;\;&
  \mbox{pairing density}
  \\[5pt]
  \hline
  \end{array}
\label{eq:rtj}
\end{equation}
In this equation,  $v_\alpha$ and $u_\alpha$ are the standard BCS (or canonical HFB)  amplitudes. The phase-space weight
$f_\alpha$
provides a smooth cutoff of the
space of single-particle states included in pairing.
All of the above expressions are local quantities that depend on the position vector $\vec{r}$ and refer
to the local wave function components $\varphi_\alpha=\varphi_\alpha(\vec{r})$.
The pairing density (\ref{eq:rtj}) is restricted to $\alpha>0$, which stands for states with positive azimuthal angular momenta (the other half with $\alpha<0$ are the pairing conjugate states). 

For the pairing cutoff, we use a soft cutoff with the profile \cite{Kri90a}
\begin{equation}
  f_\alpha
  =
  \big(1+
    \exp{\left((\varepsilon_\alpha-(\epsilon_{\mathrm{F},q_\alpha}+\epsilon_\mathrm{cut}))
            /\Delta\epsilon\right)}
\big)^{-1},
\label{eq:softcut}
\end{equation}
where $\varepsilon_\alpha$ are the single-particle energies,
 $\epsilon_\mathrm{cut}$ marks the cutoff band, and $\Delta\epsilon=\epsilon_\mathrm{cut}/10$ is its width.
We use a dynamical
setting of the pairing band where $\epsilon_\mathrm{cut}$ is adjusted such
that a fixed number of nucleons $N_q+\eta_\mathrm{cut}N_q^{2/3}$ is
included \cite{Ben00a}, here
with $\eta_\mathrm{cut}=5$ as in \cite{Reinhard2017}.

\section{Nuclear matter properties}
\label{sec:NMP}

Bulk properties of symmetric nuclear matter at equilibrium, called nuclear matter
properties (NMPs), are often used to characterize the properties of a
model, or functional respectively. A starting point for the definition
of NMPs is the binding energy per nucleon in the symmetric nuclear matter
\begin{equation}
  E_{\rm B}/A
  =
  E(\rho_0,\rho_1,\tau_0,\tau_1)/A
  =
  \frac{\mathcal{E}_{\mathrm{Fy}}^\mathrm{v}}{\rho_0}
  \;,
\label{eq:defEA}  
\end{equation}
which depends uniquely on the volume term (\ref{EFay-dens}) of the functional.
Variation with respect to Kohn--Sham  wave functions establishes a relation
$\tau_t=\tau_t(\rho_0,\rho_1)$ between kinetic densities $\tau_t$ and
densities $\rho_t$. This yields the commonly used binding energy at equilibrium, 
$E_{\rm B}/A\left[\rho_0,\rho_1,\tau_0(\rho_0,\rho_1),\tau_1(\rho_0,\rho_1)\right]$, as a function of the densities $\rho_t$ alone.
\begin{table}
\caption{\label{tab:nucmatdef}
Definitions of NMPs used in this work.
All derivatives are to be  taken at the equilibrium point corresponding to the equilibrium density $\rho_\mathrm{eq}$.
}
\begin{center}
\begin{tabular}{lrcl}
 binding energy:
 &
  $\displaystyle\frac{{E_{\rm B}}}{A}$
  &=&
    $\displaystyle
    \displaystyle\frac{{E_{\rm B}}}{A}\Big|_\mathrm{eq}$
\\[12pt]
  equilibrium density:
  &
  $\displaystyle\rho_{0,\rm{eq}}$ &$\leftrightarrow$
  &
  $\partial_{\rho_0}\frac{{E_{\rm B}}}{A}=0$
\\[12pt]
 incompressibility:
 &
  $  K_\infty$
  &=& 
  $\displaystyle
  9\,\rho_0^2 \, \frac{d^2}{d\rho_0^2} \,
       \frac{{E_{\rm B}}}{A}\Big|_\mathrm{eq}
  $
\\[12pt]
  symmetry energy:
  &
  $a_\mathrm{sym}$
  &=&
  $\displaystyle
 \frac{1}{2} \rho_0^2 \frac{d^2}{d\rho_1^2}
  \frac{{_{\rm B}}}{A} \bigg|_\mathrm{eq} \equiv J
  $
\\[12pt]
  slope of $a_\mathrm{sym}$:
  &
  $L$
  &=&
  $\displaystyle
  3\rho_0 \frac{d a_\mathrm{sym}}{d\rho_0} \bigg|_\mathrm{eq}
  $
\end{tabular}
\end{center}
\end{table}
\Tref{tab:nucmatdef} lists the NMPs discussed in this work.
We consider $\tau_t$ as independent variables for the purpose of a formally compact definition of the effective mass.  Static properties are deduced from the binding energy at equilibrium, which depends on $\rho_0$ only. This is indicated by using the total derivatives  for $K_\infty$,
$a_\mathrm{sym}$, and $L$.
The slope of the symmetry energy $L$
parameterizes
the density dependence of $a_\mathrm{sym}$.

All these NMPs depend on the volume parameters of the Fayans functional through (\ref{eq:defEA}). There are six volume parameters in
$E_{\rm B}/{A}|_\mathrm{eq}$ and five NMPs. We use the NMPs to express five of the volume parameters. 
${h_{2-}^\mathrm{v}}$ is the sole remaining  volume parameter. 

\section{Input data in detail}
\label{app:data}

\begin{table*}[bth]
\caption{\label{tab:fitdata1} Calibration data Part I: 
bulk data along   isotopic chains. \\
   }
{\small
\begin{tabular}{|rr||rr|rr|rr|rr|}
\hline
 A &    Z &
  \multicolumn{1}{|c}{$E_{\rm B}$} &
  \multicolumn{1}{c|}{$\Delta E_{\rm B}$} &
  \multicolumn{1}{|c}{$R_\mathrm{box}$} &
  \multicolumn{1}{c|}{$\Delta R_\mathrm{box}$} &
  \multicolumn{1}{|c}{$\sigma$} &
  \multicolumn{1}{c|}{$\Delta \sigma$} &
  \multicolumn{1}{|c}{$r_\mathrm{ch}$} &
  \multicolumn{1}{c|}{$\Delta r_\mathrm{ch}$} 
\\
\hline
    &     &  
  \multicolumn{2}{|c|}{MeV} &
  \multicolumn{2}{|c|}{fm} &
  \multicolumn{2}{|c|}{fm} &
  \multicolumn{2}{|c|}{fm} 
\\
\hline \hline
 16 &   8 &  -127.620 & 4 & 2.777 & 0.08 & 0.839 & 0.08 & 2.701 & 0.04\\
\hline
 36 &  20 &  -281.360 & 2 &       &   &       &   & 3.450 & 0.18  \\
 38 &  20 &  -313.122 & 2 &       &   &       &   & 3.466 & 0.10  \\
 40 &  20 &  -342.051 & 3 & 3.845 & 0.04 & 0.978 & 0.04 & 3.478 & 0.02\\
 42 &  20 &  -361.895 & 2 & 3.876 & 0.04 & 0.999 & 0.04 & 3.513 & 0.04\\
 44 &  20 &  -380.960 & 2 & 3.912 & 0.04 & 0.975 & 0.04 & 3.523 & 0.04\\
 46 &  20 &  -398.769 & 2 &       &   &       &   & 3.502 & 0.02\\
 48 &  20 &  -415.990 & 1 & 3.964 & 0.04 & 0.881 & 0.04 & 3.479 & 0.04\\
 50 &  20 &  -427.491 & 1 &       &   &       &   & 3.523 & 0.18\\
 52 &  20 &  -436.571 & 1 &       &   &       &   & 3.5531 & 0.18   \\
\hline
 58 & 26 &           &   &       &   &       &  & 3.7745 & 0.18\\
\hline
 56 &  28 &  -483.990 & 5 &       &   &       &   & 3.750 & 0.18\\
 58 &  28 &  -506.500 & 5 & 4.364 & 0.04 &       &   & 3.776 & 0.10\\
 60 &  28 &  -526.842 & 5 & 4.396 & 0.04 & 0.926 & 0.20 & 3.818 & 0.10\\
 62 &  28 &  -545.258 & 5 & 4.438 & 0.04 & 0.937 & 0.20 & 3.848 & 0.10\\
 64 &  28 &  -561.755 & 5 & 4.486 & 0.04 & 0.916 & 0.08 & 3.868 & 0.10\\
 68 &  28 &  -590.430 & 1 &       &   &       &   &  &    \\
\hline
100 &  50 &  -825.800 & 2 &       &   &       &   &       &    \\
108 &  50 &           &   &       &   &       &   & 4.563 & 0.04\\
112 &  50 &           &   & 5.477 & 0.12 & 0.963 & 0.36 & 4.596 & 0.18\\
114 &  50 &           &   & 5.509 & 0.12 & 0.948 & 0.36 & 4.610 & 0.18\\
116 &  50 &           &   & 5.541 & 0.12 & 0.945 & 0.36 & 4.626 & 0.18\\
118 &  50 &           &   & 5.571 & 0.08 & 0.931 & 0.08 & 4.640 & 0.02\\
120 &  50 &           &   & 5.591 & 0.04 &       &      & 4.652 & 0.02\\
122 &  50 & -1035.530 & 3 & 5.628 & 0.04 & 0.895 & 0.04 & 4.663 & 0.02\\
124 &  50 & -1050.000 & 3 & 5.640 & 0.04 & 0.908 & 0.04 & 4.674 & 0.02\\
126 &  50 & -1063.890 & 2 &       &   &       &   &  &   \\
128 &  50 & -1077.350 & 2 &       &   &       &   &  &   \\
130 &  50 & -1090.400 & 1 &       &   &       &   &  &   \\
132 &  50 & -1102.900 & 1 &       &   &       &   &  &   \\
134 &  50 & -1109.080 & 1 &       &   &       &   &  &   \\
\hline
198 &  82 & -1560.020 & 9 &       &   &       &   & 5.450 & 0.04\\
200 &  82 & -1576.370 & 9 &       &   &       &   & 5.459 & 0.02\\
202 &  82 & -1592.203 & 9 &       &   &       &   & 5.474 & 0.02\\
204 &  82 & -1607.521 & 2 & 6.749 & 0.04 & 0.918 & 0.04 & 5.483 & 0.02\\
206 &  82 & -1622.340 & 1 & 6.766 & 0.04 & 0.921 & 0.04 & 5.494 & 0.02\\
208 &  82 & -1636.446 & 1 & 6.776 & 0.04 & 0.913 & 0.04 & 5.504 & 0.02\\
210 &  82 & -1645.567 & 1 &       &   &       &   & 5.523 & 0.02\\
212 &  82 & -1654.525 & 1 &       &   &       &   & 5.542 & 0.02\\
214 &  82 & -1663.299 & 1 &       &   &       &   & 5.559 & 0.02\\
\hline
\end{tabular}
}
\end{table*}

\begin{table*}
\caption{\label{tab:fitdata2}
 Calibration data Part II. Similar as in \tref{tab:fitdata1},  but for nuclei along isotonic chains.\\
}
\begin{tabular}{|rr||rr|rr|rr|rr|}
\hline
 A &    Z &
  \multicolumn{1}{|c}{$E_{\rm B}$} &
  \multicolumn{1}{c|}{$\Delta E_{\rm B}$} &
  \multicolumn{1}{|c}{$R_\mathrm{box}$} &
  \multicolumn{1}{c|}{$\Delta R_\mathrm{box}$} &
  \multicolumn{1}{|c}{$\sigma$} &
  \multicolumn{1}{c|}{$\Delta \sigma$} &
  \multicolumn{1}{|c}{$r_\mathrm{ch}$} &
  \multicolumn{1}{c|}{$\Delta r_\mathrm{ch}$} 
\\
\hline
    &     &  
  \multicolumn{2}{|c|}{MeV} &
  \multicolumn{2}{|c|}{fm} &
  \multicolumn{2}{|c|}{fm} &
  \multicolumn{2}{|c|}{fm} 
\\
\hline \hline
 34 &  14 &  -283.429 & 2  &       &   &       &   &       &    \\
 36 &  16 &  -308.714 & 2  & 3.577 & 0.16 & 0.994 & 0.16 & 3.299 & 0.02\\
 38 &  18 &  -327.343 & 2  &       &   &       &   & 3.404 & 0.02\\
\hline
 50 &  22 &  -437.780 & 2  & 4.051 & 0.04 & 0.947 & 0.08 & 3.570 & 0.02\\
 52 &  24 &   &  & 4.173 & 0.04 & 0.924 & 0.16 & 3.642 & 0.04\\
 54 &  26 &   &  & 4.258 & 0.04 & 0.900 & 0.16 & 3.693 & 0.04\\
\hline
 86 &  36 &  -749.235 & 2 &       &   &       &   & 4.184 & 0.02 \\
 88 &  38 &  -768.467 & 1 & 4.994 & 0.04 & 0.923 & 0.04 & 4.220 & 0.02 \\
 90 &  40 &  -783.893 & 1 & 5.040 & 0.04 & 0.957 & 0.04 & 4.269 & 0.02\\
 92 &  42 &  -796.508 & 1 & 5.104 & 0.04 & 0.950 & 0.04 & 4.315 & 0.02\\
 94 &  44 &  -806.849 & 2 &       &   &       &   &       &   \\
 96 &  46 &  -815.034 & 2 &       &   &       &   &       &   \\
 98 &  48 &  -821.064 & 2 &       &   &       &   &       &   \\
\hline
134 &  52 & -1123.270 & 1 &       &   &       &   &       &   \\
136 &  54 & -1141.880 & 1 &       &   &       &   & 4.791 & 0.02\\
138 &  56 & -1158.300 & 1 & 5.868 & 0.08 & 0.900 & 0.08 & 4.834 & 0.02\\
140 &  58 & -1172.700 & 1 &       &   &       &   & 4.877 & 0.02\\
142 &  60 & -1185.150 & 2 & 5.876 & 0.12 & 0.989 & 0.12 & 4.915 & 0.02\\
144 &  62 & -1195.740 & 2 &       &   &       &   & 4.960 & 0.02\\
146 &  64 & -1204.440 & 2 &       &   &       &   & 4.984 & 0.02\\
148 &  66 & -1210.750 & 2 &       &   &       &   & 5.046 & 0.04\\
150 &  68 & -1215.330 & 2 &       &   &       &   & 5.076 & 0.04\\
152 &  70 & -1218.390 & 2 &       &   &       &   &       &   \\
\hline
206 &  80 & -1621.060 & 1 &       &   &       &   & 5.485 & 0.02\\
210 &  84 & -1645.230 & 1 &       &   &       &   & 5.534 & 0.02\\
212 &  86 & -1652.510 & 1 &       &   &       &   & 5.555 & 0.02\\
214 &  88 & -1658.330 & 1 &       &   &       &   & 5.571 & 0.02\\
216 &  90 & -1662.700 & 1 &       &   &       &   &       &   \\
218 &  92 & -1665.650 & 1 &       &   &       &   &       &   \\
\hline
\end{tabular}
\end{table*}

\begin{table*}[tbp]
\caption{\label{tab:fitlsgap}
Calibration data Part III: spin-orbit splittings (upper block)
and adopted errors of  3-point binding energy differences (lower block) for
  neutrons $\Delta^{(3)}_n E(Z,N)
  =\frac{1}{2}(E_{\rm B}(Z,N\!+\!2)-2E_{\rm B}(Z,N)+E_{\rm B}(Z,N\!-\!2))$ and
  for protons
  $\Delta^{(3)}_p E(Z,N)
  =\frac{1}{2}(E_{\rm B}(Z\!+\!2,N)-2E_{\rm B}(Z,N)+E_{\rm B}(Z\!-\!2,N))$. All quantities are in MeV.  }
\begin{center}
\begin{tabular}{|cc||ccc|ccc|}
\hline
 $A$ &    $Z$ &
  Level&
  \multicolumn{1}{c}{$\varepsilon_{ls,p}$} &
  \multicolumn{1}{c|}{$\Delta\varepsilon_{ls,p}$} &
  Level&
  \multicolumn{1}{c}{$\varepsilon_{ls,n}$} &
  \multicolumn{1}{c|}{$\Delta\varepsilon_{ls,n}$} 
\\
\hline 
 16 &   8 & 1p & 6.30 &  60\%   & 1p & 6.10 &  60\% \\
132 &  50 & 2p & 1.35 &  20\%   & 2d & 1.65 &  20\% \\
208 &  82 & 2d & 1.42 &  20\%   & 1f & 0.90 &  20\% \\
    &     &    &      &         & 3p &  1.77 & 40\% \\
\hline
\end{tabular}
\\[10pt]
\begin{tabular}{|rr|cc||rr|cc|}
\hline
\multicolumn{4}{|c||}{$\Delta^{(3)}_n E$ } 
&
\multicolumn{4}{|c|}{$\Delta^{(3)}_p E$ } 
\\
  \hline 
 $A$ & $Z$ & Data & Error  &  $A$ & $Z$ & Data & Error  \\
\hline
 44 & 20 & 0.628 & 0.24 &  36& 16 & 3.328 & 0.36\\
118 & 50 & 0.330 & 0.36 &  88& 38 & 1.903 & 0.36\\
120 & 50 & 0.300 & 0.36 &  90& 40 & 1.4055 & 0.24\\
122 & 50 & 0.260 & 0.24 &  92& 42 & 1.137 & 0.12\\
124 & 50 & 0.290 & 0.24 &  94& 44 & 1.078 & 0.24\\
&&&& 136& 54 & 1.095 & 0.24 \\
&&&& 138& 56 & 1.010 & 0.24 \\
&&&& 140& 58 & 0.975 & 0.24 \\
&&&& 142& 60 & 0.930 & 0.24 \\
&&&& 214& 88 & 0.725 & 0.24 \\
&&&& 216& 90 & 0.710 & 0.24 \\
\hline 
\end{tabular}
\end{center}
\end{table*}

\begin{table}[bh]
\caption{\label{tab:new-iso}
Calibration data Part IV:  differential charge  radii
$\delta \langle r^2\rangle^{A,A'} = \langle r^2_{\rm ch} \rangle^{A} -
 \langle r^2_{\rm ch} \rangle^{A'}$ (in fm$^2$).
}
\medskip
\centerline{
\begin{tabular}{|rrr|rr|}
\hline
\multicolumn{5}{|c|}{$\delta \langle r^2\rangle^{A,A'}\;$ } 
\\
\hline 
 A & A' & Z & Data & Error \\
\hline
 48 & 40 & 20 & 0.006957 & 0.008 \\
 48 & 44 & 20 & -0.308088 & 0.008 \\
 52 & 48 & 20 & 0.52107861 & 0.020 \\
\hline
\end{tabular}
}
\end{table}
Tables~\ref{tab:fitdata1}, \ref{tab:fitdata2}, \ref{tab:fitlsgap}, and \ref{tab:new-iso} show the detailed calibration data $\cal D$  together with
their adopted error.

\clearpage

\section*{References}
\bibliography{add_abbrev}
\bibliographystyle{iopart-num}

\clearpage
\vspace{3em}

\small

\framebox{\parbox{\linewidth}{
The submitted manuscript has been created by UChicago Argonne, LLC, Operator of 
Argonne National Laboratory (``Argonne''). Argonne, a U.S.\ Department of 
Energy Office of Science laboratory, is operated under Contract No.\ 
DE-AC02-06CH11357. 
The U.S.\ Government retains for itself, and others acting on its behalf, a 
paid-up nonexclusive, irrevocable worldwide license in said article to 
reproduce, prepare derivative works, distribute copies to the public, and 
perform publicly and display publicly, by or on behalf of the Government.  The 
Department of Energy will provide public access to these results of federally 
sponsored research in accordance with the DOE Public Access Plan. 
\url{http://energy.gov/downloads/doe-public-access-plan.}}}
	
\end{document}